\documentclass[pdftex,a4paper]{scrartcl}

\usepackage{abstract}
\usepackage{color}
\usepackage{comment}
\usepackage[pdftex]{graphicx}
\usepackage[utf8]{inputenc}
\usepackage{natbib}
\usepackage{tabularx}
\usepackage{tabulary}
\usepackage{array}%
\usepackage{relsize}
\usepackage{setspace}
\usepackage{booktabs, makecell}
\usepackage{siunitx}
\usepackage{authblk}
\usepackage{lineno}

\usepackage{subfiles}
\usepackage{subcaption}
\usepackage{hyperref}

\newcommand{\mj}[1]{\textcolor{green}{/* #1 (mj) */}}

\title{Individual and gender inequality in computer science: A career study of cohorts from 1970 to 2000}

\author[1]{Haiko Lietz}
\author[1,2]{Mohsen Jadidi}
\author[1]{Daniel Kostic}
\author[3]{Milena Tsvetkova}
\author[1,4]{Claudia Wagner}
\affil[1]{GESIS -- Leibniz Institute for the Social Sciences, Cologne, Germany}
\affil[2]{University of Koblenz-Landau, Koblenz, Germany}
\affil[3]{The London School of Economics and Political Science, London, UK}
\affil[4]{RWTH Aachen University, Aachen, Germany}

\begin{document}

\maketitle

\begin{abstract}
\noindent Inequality prevails in science. Individual inequality means that most perish quickly and only a few are successful, while gender inequality implies that there are differences in achievements for women and men. Using large-scale bibliographic data and following a computational approach, we study the evolution of individual and gender inequality for cohorts from 1970 to 2000 in the whole field of computer science as it grows and becomes a team-based science. We find that individual inequality in productivity (publications) increases over a scholar's career but is historically invariant, while individual inequality in impact (citations), albeit larger, is stable across cohorts and careers. Gender inequality prevails regarding productivity, but there is no evidence for differences in impact. The Matthew Effect is shown to accumulate advantages to early achievements and to become stronger over the decades, indicating the rise of a ``publish or perish'' imperative. Only some authors manage to reap the benefits that publishing in teams promises. The Matthew Effect then amplifies initial differences and propagates the gender gap. Women continue to fall behind because they continue to be at a higher risk of dropping out for reasons that have nothing to do with early-career achievements or social support. Our findings suggest that mentoring programs for women to improve their social-networking skills can help to reduce gender inequality.

\bigskip

\end{abstract}

\section{Introduction}

Half a century ago, \citeauthor{price1963little} diagnosed that the science system exhibits an ``essential, built-in undemocracy,'' meaning that academic achievements are strongly concentrated among a very limited number of persons or organizations. He observed inequality in the form of broad distributions of individual productivity and scientific impact and found this pattern to be stable as science grows, perpetuating a system where a ``few giants'' coexist with a ``mass of pygmies'' \citep[p.53]{price1963little}.
The literature has found the broadness of these distributions to be a universal property of the science system \citep{Lotka1926, bradford_sources_1934, albarran_skewness_2011, RUIZCASTILLO2014917} and has identified an endogenous process of reproduction as the main driving mechanism: the Matthew Effect \citep{diprete_cumulative_2006, perc_matthew_2014, bol_matthew_2018}.
In his explanations of advancement in academic careers, \citet{merton_matthew_1968, merton_matthew_1988} referred to the Matthew Effect (ME) as a cumulative-advantage process according to which ``initial comparative advantages of trained capacity, structural location, and available resources make for successive increments of advantage such that the gaps between the haves and the have-nots in science ... widen until dampened by countervailing processes.'' \citep[p.~606]{merton_matthew_1988} The larger the ME, the more ``the rich get richer rendering the poor relatively poorer'' \citep[p.~34]{page_what_2015}.

Extreme individual inequality is problematic but could be considered fair if it is merit-based \citep{starmans_why_2017}. However, for differences in merit and success to be considered fair, they should not be associated with ascribed characteristics such as gender, age, or ethnicity \citep{merton_normative_1973, cole_fair_1979}. Inequality among persons belonging to different groups, also known as horizontal inequality as opposed to individual, or vertical, inequality, is undesirable \citep{stewart2005horizontal}.
For example, gender is a prominent principle of distinction, and gender inequality in scientific productivity has been observed. This is known as the ``productivity puzzle.'' For instance, research from the early days of science studies found that women produce about half as much as men \citep{Cole1984, cole_theory_1991}, particularly over the first decade of their careers \citep{reskin_scientific_1979, Long1992}. More recent large-scale analyses show that each year, women are 20 percent more likely to drop out of science than men \citep{huang_historical_2020}. In computer science, women on average publish less than men per year for the first several years of employment \citep{Way2016}. Women are less likely to take prestigious author positions in publications \citep{West2013, holman_gender_2018}, yet they are more likely to perform better in the job market \citep{Way2016}.
Gender inequality in impact has also been reported \citep{Cole1984, lincoln_matilda_2012, Cassidy2013}.


The literature on individual and gender inequality in science is abundant, but we identify two major research gaps in it. The first relates to cohort design and data availability. Older analyses tend to have sound cohort designs but are often restricted in the amount of data (number and size of cohorts) that were studied. For example, \citet{zuckerman_age_1972} only analyze one cohort, while \citet{Allison1982} analyze three cohorts. More recent computational analyses tend to study large amounts of data but are often restricted regarding cohort design. For example, \citet{penner_predictability_2013} aggregated scientists that started their careers in the same decade. \citet{petersen2014reputation} group authors into one cohort that published their first paper in a competitive journal within the same 15 years. These cohorts are heterogeneous with respect to career age and do not include unsuccessful scientists and early career researchers. Previous research, however, has shown that life-course approaches are important because dropouts can partially explain gender inequality. Specifically, productivity inequality almost vanishes when women and men are compared for the same career ages \citep{Jadidi2017, azoulay_self_citation_2020, huang_historical_2020}, although, even when the survival bias is removed, women still have fewer publications than men when they become a professor \citep{aksnes_are_2011, lutter_who_2016}. 

The second research gap relates to recognizing the scientific field's growth and transformation.  
\citet{price1963little} argued that recruiting more people into science implies that less talented people will enter. \citet{zuckerman_age_1972} hypothesized that this leads to larger differences between the most and the least talented, suggesting that inequality should be higher in more recent cohorts than in older ones. Early work on chemistry cohorts found inequality in productivity (publications) and scientific impact (citations) to increase as a cohort ages \citep{Allison1982}. Yet, using full-scale bibliographic databases, scholars found impact inequality decreases over time \citep{Lariviere2009, Petersen2014, Pan2018} as the academic system transitions from a scholar-centered to a globalized, interdisciplinary, team- and project-based mode of knowledge production \citep{gibbons_new_1994}. Both findings are plausible and can be explained by changes in the academic system: the increased tendency to publish papers with multiple authors \citep{Wuchty2007, Petersen2014Teams} may function as a social multiplier that potentially increases inequality, while a higher number of references per paper decreases the number of uncited papers which may decrease inequality \citep{Wallace2009, Pan2018}.

In this paper, we take a cohort-design approach to the problem of individual and gender inequalities in academia and their origins. Using bibliographic data on the whole field of computer science, we define cohorts from 1970 to 2000 and study the careers of authors over 15 years. Computer science presents an ideal case study because we can observe it since its early days as it grows and evolves from an individual-based to a team-based science. The field is relatively young, growing, in ongoing transformation, and a driver of the digital revolution. Last but not least, it concerns potentially large gender disparities since only one out of five computer scientists is female \citep{lee_homophily_2019}.
We find that individual inequality in productivity is slightly increasing over academic careers. In contrast, individual inequality in impact is stable. These trends are invariant as the field grows and matures. Gender inequality exists, but impact inequality finds an explanation in productivity inequality which is a result of higher dropout rates for women. The ME is shown to increase historically. Over the decades, we expose the emergence of an imperative to ``publish or perish'' and the citation-based consequences of early-career achievements as well as early-career social capital. By shedding light on the mechanisms behind individual and gender inequality we motivate science policy interventions to mentor women in social networking. 


In the next section, we distill from the literature an evolutionary theory of careers in competitive fields with the ME at its center. This theory guides our analysis. Then, we present our research design, discuss our results in detail, and conclude our work. For readability, materials and methods are placed at the end.

\section{The Matthew Effect in the center of a theory of careers}

The ME is a feedback mechanism that generates inequality. On the one hand, the ME implies \textit{cumulative advantage}.
For example, getting a more prestigious job entails an increase in productivity \citep{allison_departmental_1990}. 
Departmental prestige helps careers because prestige operates and reproduces in networks. As a scholar climbs up the career ladder, she or he advances into the core of a field and becomes part of a reproductive vortex that makes it increasingly hard to \textit{not} benefit from collective dynamics \citep{burris_academic_2004, clauset_systematic_2015, way_productivity_2019}. Cores harbor the few positions that strongly influence how a field reproduces \citep{fuchs_against_2001}. \citet{padgett_emergence_2012} introduce the concept of autocatalytic feedback to model these dynamics.

Inversely, the ME also takes the form of a \textit{cumulative disadvantage}. This has sustained the hypothesis that success either comes early or not at all \citep{zuckerman_age_1972}. 
As a consequence, the ME makes it increasingly difficult for an individual to stay in academia \citep{cole1973social}. Young scientists must overcome a ``barrier'' to excel \citep{PetersenPNAS2011}. If positive feedback does not set in early in a career, the respective scholar requires motivation to be productive for the love of the work or some amount of tenacity \citep{huber_new_2002}. Surprisingly, though most computer scientists are most productive in their fifth year after hiring, there is a huge variance in productivity career patterns \citep{way_misleading_2017}. And success can come at any time in a career, but it depends on persistence, ability to excel, and, last but not least, luck \citep{sinatra_quantifying_2016}.

Cumulative advantage and disadvantage both imply that past achievement to some extent predicts current achievement. Thus, empirical research on the ME typically quantifies the size of the effect and even attempts to establish a scaling law \citep{jeong_measuring_2003, perc_matthew_2014, rondapupo_evolutions_2018}.
Career reinforcement via the ME manifests as increasing returns to the average number of citations per paper as an author becomes more productive \citep{costas_scaling_2009}. For highly-cited authors, staying in academia twice as long means being up to $2.8$ times more productive and being up to eight times more impactful. Below a certain citation threshold, the ME operates via the author's reputation as measured by their cumulative citation record, but above that threshold, mainly via publication visibility \citep{petersen2014reputation}. 
Overall, studies predicting the success of scholars or publications have found that current productivity and impact \citep{acuna_future_2012, penner_predictability_2013, mazloumian_predicting_2012, Dong2015}, combined with an intrinsic ``fitness'', or ability and quality \citep{wang_quantifying_2013}, and mediated through networks \citep{sarigol_predicting_2014} are positively correlated with future success.
The observation that the early career of a scientist is predictive of her or his later success and gains in predictive power diminish as more career ages are used for prediction provides further evidence for the ME \citep{mazloumian_predicting_2012, penner_predictability_2013, wang_quantifying_2013}.

In sum, the ME has become central to an empirically-oriented evolutionary \textit{theory of careers} in competitive fields that is taking shape at the intersection of the social and computational sciences.
It is a \textit{field theory} \citep{bourdieu_homo_1988} because the academic fields, as spacetimes that delimit agents' social positions and interactions, are the loci that harbor the ME \citep{white_networks_2004}. Emerging from collective action, field structure acts as a memory in which advantages accumulate and lead to institutionalization \citep{Petersen2014, flack_coarsegraining_2017, Pan2018}. This field-endogenous feedback process operates behind (i.e., it reinforces or impedes) life-course factors such as creativity, self-perceptions, dispositions, access to resources, and environmental conditions \citep{cole1973social, cole_theory_1991, padgett_emergence_2012}.
\textit{Competition} for ideas, positions, and funds ensues. Careers are tournament-like endeavors \citep{sorensen_social_1986} to improve one's rank in the academic ``pecking order'' \citep{chase_social_1980}. Ranks translate to positions in networks, and upward or downward mobility resembles approaching or withdrawing from network cores \citep{burris_academic_2004, clauset_systematic_2015}. Only a few make it up those ``chains of opportunity,'' for most the way is down \citep{white_chains_1970}.
As an \textit{evolutionary theory}, it looks for path dependence and the long-term consequences of initial conditions \citep{cole_theory_1991, wray_kuhns_2011}. Small differences in ability, persistence, or luck accumulate and lock a career into an upward or downward path \citep{petersen2012persistence, way_productivity_2019}. Since this is a collective phenomenon, good ideas can fail if they are put forth at the ``wrong time'' \citep{newman_first-mover_2009, bornholdt_emergence_2011}, but if the time is ``right,'' success breeds success in an avalanche-like way \citep{mazloumian_how_2011}.

This theory also prepares the ground for understanding gender inequality as co-generated by the ME \citep{long_scientific_1995, diprete_cumulative_2006}.
Some or many of the career factors exemplified above are likely to be gender-correlated and thus generate outcome differences that increase over the career as they interact with the ME \citep{Xie1998, cole_theory_1991}.
For example, absence from the job market (e.g., because of motherhood) leads to disadvantages that accumulate \citep{cole_fair_1979, diprete_cumulative_2006}. And women's disadvantages grow early in a career \citep{reskin_scientific_1979, Long1992}. 

\section{Research design}

We adopt an integrated modeling approach to study individual and gender inequality in an academic field. By ``integrated'' we mean that we are interested in both explaining and predicting inequality \citep{hofman_integrating_2021}.
We study 15-year careers in the entire computer science discipline for cohorts from 1970 to 2000.
Starting with descriptive modeling, we explore the evolution of individual and gender inequality in productivity and impact over the career within cohorts and between cohorts over time.
In an explanatory modeling step, we then present the ME as a plausible mechanism that generates the patterns of individual inequality we observe.
Finally, in a predictive modeling step, we inquire how accurately the early career predicts total-career achievements.
We identify the meritocratic and non-meritocratic early-career factors that predict whether an author drops out of the field and how successful they become eventually. Explanations of individual and gender inequality then derive from the assumption that the ME accumulates the early advantages from these career factors.

\begin{figure*}
    \centering
    \includegraphics[scale=0.28]{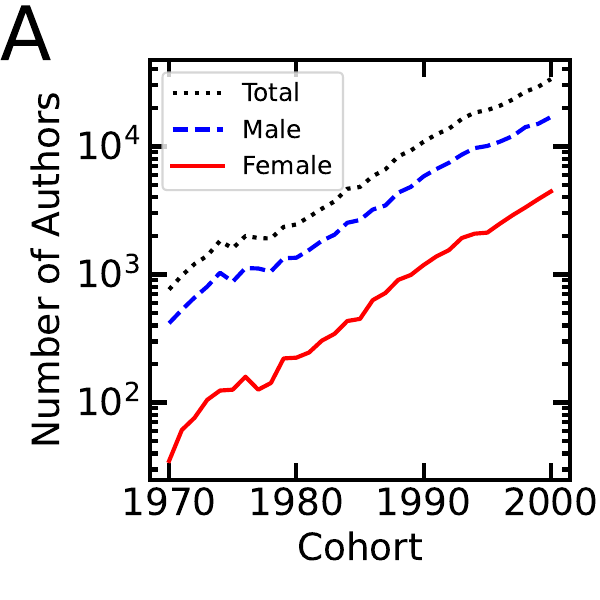}
    \includegraphics[scale=0.28]{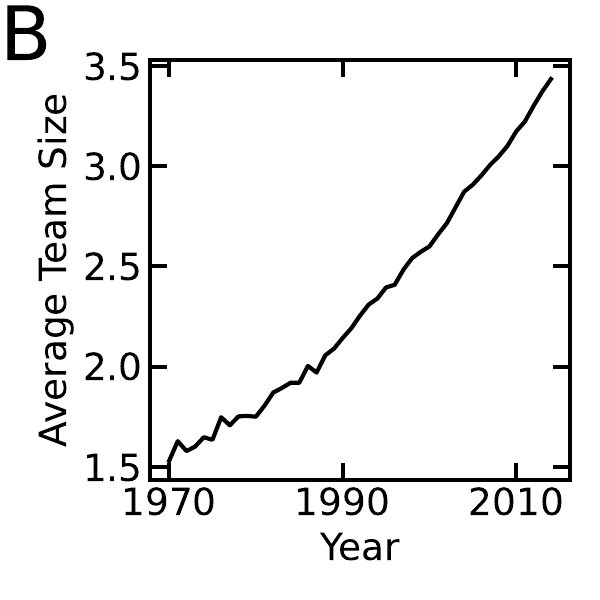}
    \includegraphics[scale=0.28]{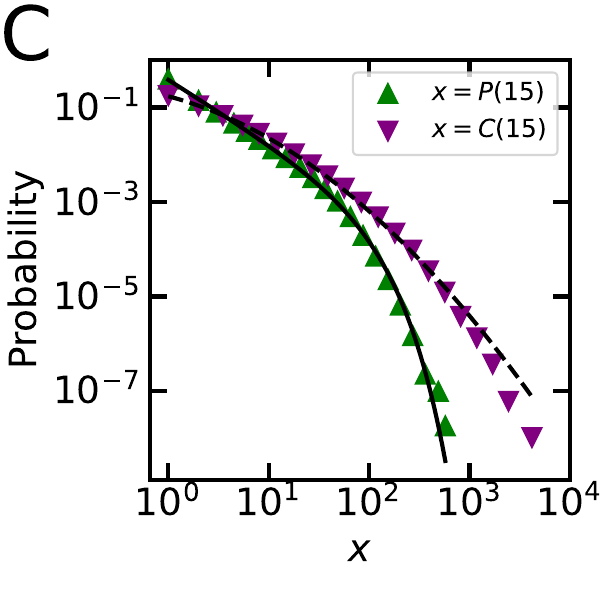}
    \includegraphics[scale=0.28]{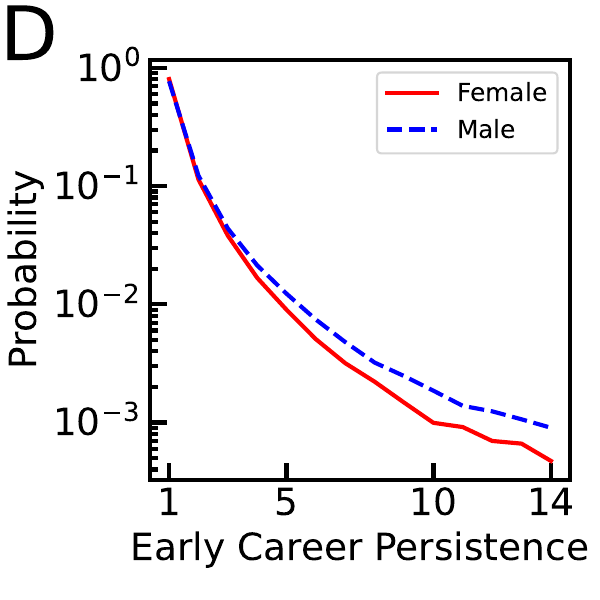}
    \includegraphics[scale=0.28]{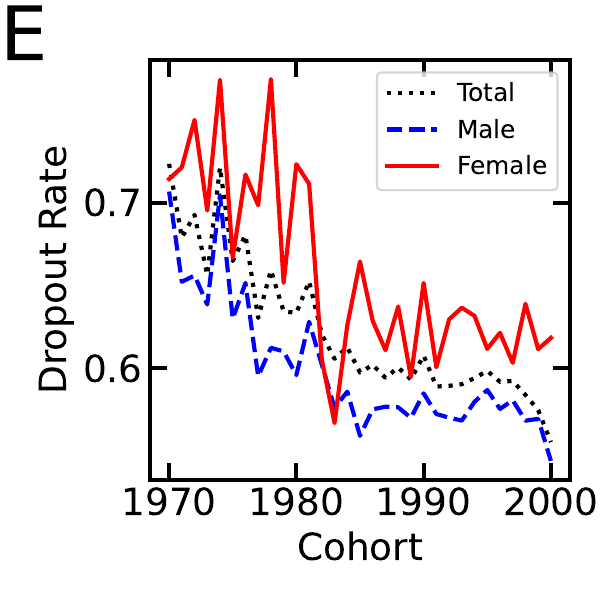}
    \caption{Description of the field of computer science. (A) The size of cohorts increases exponentially with time for both males and females. (B) The average team size, measured by the number of authors per paper, increases over time. (C) Distributions of productivity (cumulative number of papers $P$ per author at career age 15) and impact (cumulative number of citations $C$ per author at career age 15) are broad. The lines are best fits to the data: a truncated power law ($P(15)$) and a stretched exponential ($C(15)$). (D) The number of authors decreases with the number of years during which they publish persistently after the beginning of their careers (early career persistence). Female scientists show equal persistence in their early careers but after 4 years they are less likely to persist. (E) The fraction of authors in a cohort that drop out of academia (for ten years in a row) decreases but is more or less constant since the mid-80s. Females drop out more than men.
    }
    \label{fig_dataset_descriptives}
\end{figure*}

As the main data source, we use DBLP, a comprehensive collection of computer science papers that were published in major and minor computer science outlets \citep{DBLP:journals/pvldb/Ley09}. We study cohorts from 1970 to 2000, where an author belongs to a cohort if they have published their first paper in the given year. For each cohort, we study careers over 15 years, including the start year. We measure productivity in terms of the number of publications since those are the vehicles of academic communication \citep{merton_matthew_1968} and scientific impact in terms of the number of citations, a widely used measure \citep{merton_matthew_1988, aksnes_2019}. For details of our methods, we refer to the ``Materials and methods'' section at the end of the paper.
Selected results obtained from the DBLP dataset \citep{Way2016, Jadidi2017} were reported above in the Introduction.
Our cut of the DBLP dataset consists of 2.5 million 
publications from 1970 to 2014 that are authored by 1.4 million 
authors. Of those, about 300,000 authors
started their careers between 1970 and 2000 and are counted as cohort members. There are 7.9 million
citations among the authors' publications which we use for the impact analyses.

Figures \ref{fig_dataset_descriptives}A and \ref{fig_dataset_descriptives}B show that cohorts grow exponentially with time and that the field is becoming a team science in the process. Individual inequality at the most aggregate level (all publications and citations accumulated over an author's 15-year career, aggregated for all cohorts) is depicted via broad probability distributions. The citation distribution is broader than the productivity distribution, that is, inequality in impact is larger than inequality in productivity (figure \ref{fig_dataset_descriptives}C).
Correspondingly, the Gini coefficient, our measure of individual inequality, is larger for impact ($0.83$) than for productivity ($0.68$). This is not surprising since authors are physically constrained about the number of projects they can work on during any year but there are no such restrictions when it comes to the number of citations their work receives. 
The last two plots show early career \textit{persistence}, that is, the number of career years during which an author publishes consecutively from the beginning of their career, and the dropout rate per cohort. Most authors persist for only one year before they become inactive (for at least a year) or drop out of computer science. Long persistence is decreasingly likely, especially for female scientists (figure \ref{fig_dataset_descriptives}D). Dropout rates decrease for subsequent cohorts but women continue to be more likely to drop out than men (figure \ref{fig_dataset_descriptives}E).  

\section{Results}
\subsection{Individual inequality over careers and cohorts}
\label{sec_vertical_time}

In the first, \emph{descriptive} modeling step, we explore the evolution of individual and gender inequality regarding productivity and impact.
If the ME is in place, how would inequality change over the career?
Intuitively, one might expect that inequality should increase if the rich get richer and that an increase in productivity inequality should directly translate to an increase in impact inequality.
This is what \citet{Allison1982} find in their aforementioned study of the chemistry cohorts from the 50s and 60s. But they also find that the method of counting publications and citations -- window vs. cumulative counting -- is decisive. They find stable impact inequality for cumulative counting; increases are found only for window counting. Here, we report results using cumulative counting but also include plots with 3-year window counting in Appendix \ref{appendix}. Our measure of individual inequality is the Gini coefficient. 

\begin{figure*}[t!]
    \centering
    All authors\\
    Every Author assignment {} {} {} {} First Author assignment\\
    \includegraphics[scale=0.28]{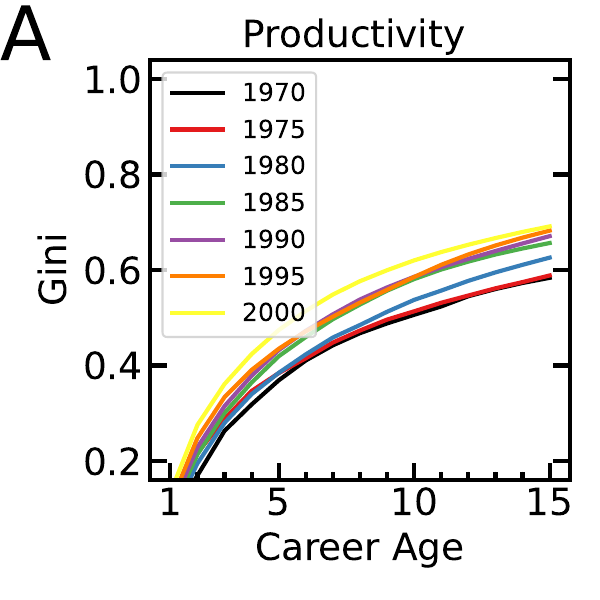}
    \includegraphics[scale=0.28]{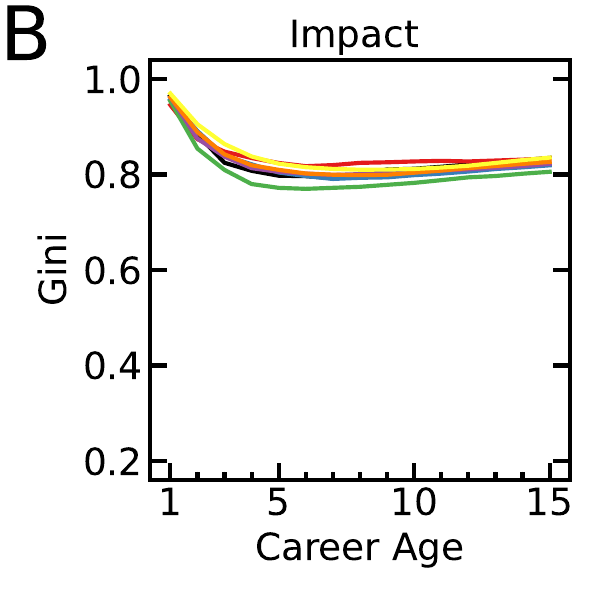}  
    \includegraphics[scale=0.28]{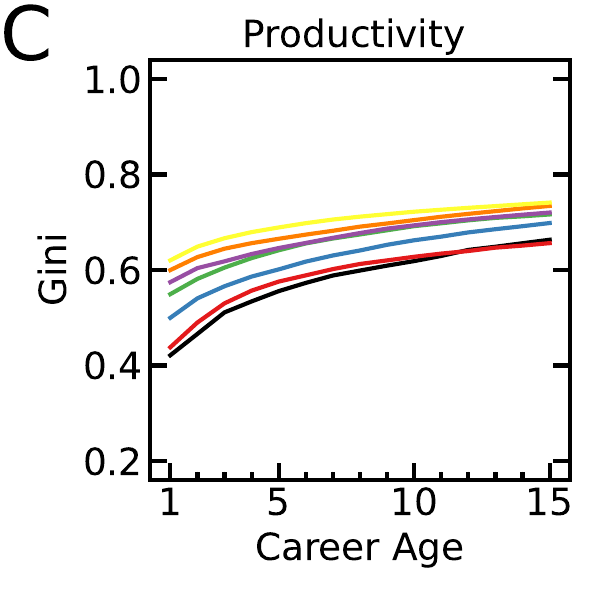}              
    \includegraphics[scale=0.28]{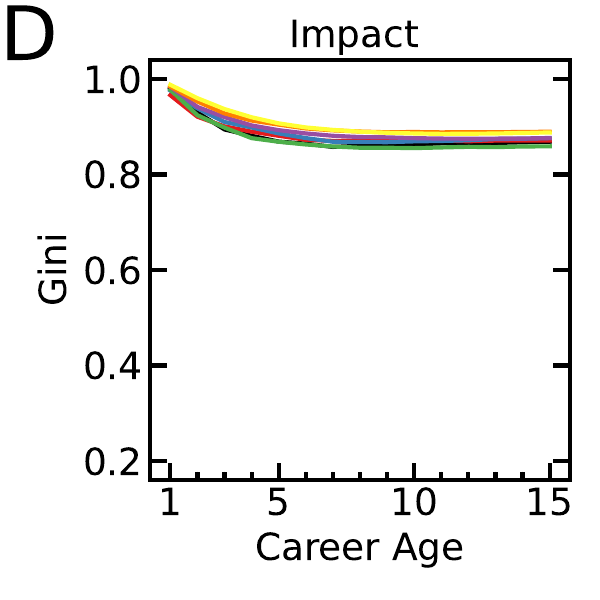}\\
    Dropouts removed\\
    Every Author assignment {} {} {} {} First Author assignment\\
    \includegraphics[scale=0.28]{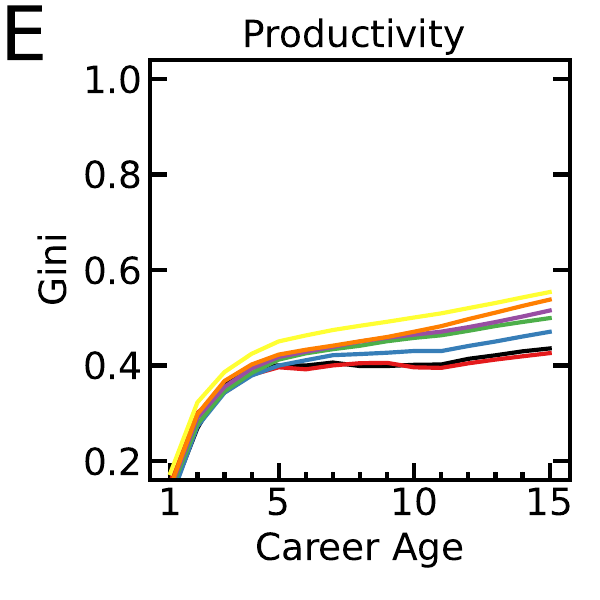}
    \includegraphics[scale=0.28]{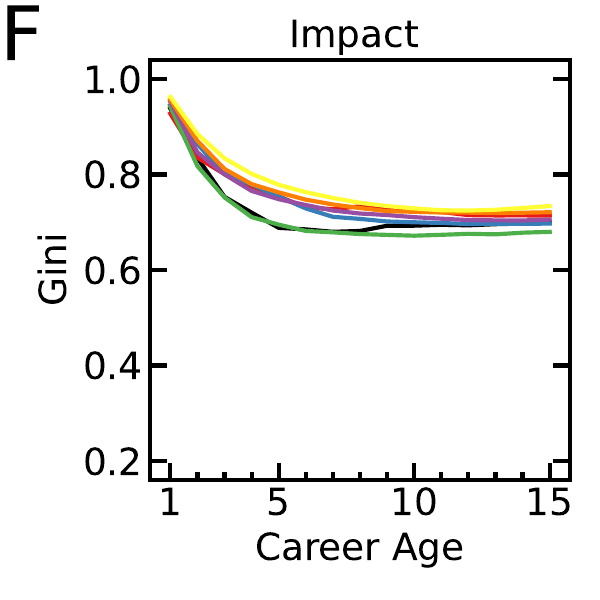}  
    \includegraphics[scale=0.28]{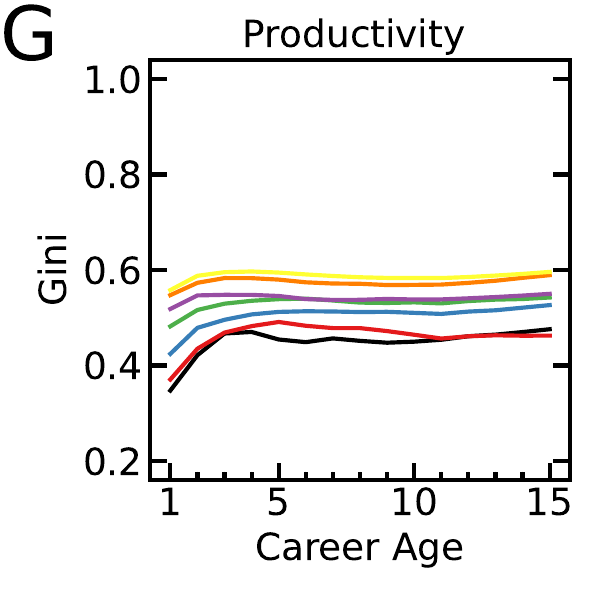}
    \includegraphics[scale=0.28]{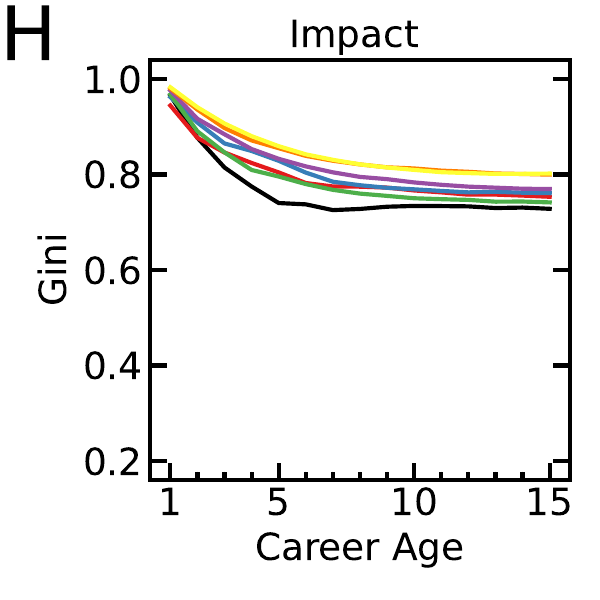}
    \caption{Individual inequality in productivity and impact as a function of career ages, depicted for seven cohorts between 1970 and 2000.
    We count publications and citations cumulatively ($P(t)$ and $C(t)$, defined in ``Materials and methods: Individual inequality'').
    (First two columns) Assigning publications to all authors. (Last two columns) Assigning publications only to first authors.
    (Second row) Authors are filtered that have not published for ten consecutive years (most likely left academia).
    }
    \label{fig:gini-cum}
\end{figure*}

For cumulative counting, we find that productivity inequality is increasing over career years (figure \ref{fig:gini-cum}A) while impact inequality is larger but mostly stable after an initial decrease (\ref{fig:gini-cum}B).
We study several modifications to validate this finding.
The change around career year 4 that can be seen in almost all figures is because the career of an author starts with the first publication (i.e., in career year 1 every author has at least one publication while even those authors that eventually become highly cited may still have zero citations).
As we saw in figure \ref{fig_dataset_descriptives}E, many authors drop out of academia early on, but their publication and citation counts influence the Gini coefficients.
We introduce the convention that we observe a \textit{dropout} if an author is absent for at least ten consecutive years.
When we remove dropouts\footnote{Results are qualitatively similar for absences of five and ten consecutive years.} (figure \ref{fig:gini-cum}E),
then author careers are more comparable and productivity inequality drops, but a trend of increasing inequality remains.
When this filter is applied to measuring impact, the inequality level also drops but the trend toward stable inequality does not change (figure \ref{fig:gini-cum}F).

In computer science, the order of authors is typically important. The first author usually did the most valued part of the work. Hence, in our analysis, attributing publications only to first authors serves the purpose of studying scholars of heightened importance.\footnote{In our cut of the DBLP dataset, 69\% of all publications have author lists that are not alphabetically sorted. Since an author ranking by importance can be alphabetic by chance, the fraction where the author ranking is indicative of importance will be even higher.}
When we add the first-author filter to the removal of dropouts, the trend for increasing productivity inequality completely vanishes (\ref{fig:gini-cum}G), and the trend for stable impact inequality remains (\ref{fig:gini-cum}H).
When we employ window counting, the Gini coefficients are systematically higher and the trends less pronounced due to the ceiling effect, but qualitatively almost the same. A difference is, though, that productivity inequality still increases slightly over a career when dropout and first-author filters are applied (figure \ref{fig:gini-win} in appendix \ref{appendix}).
In sum, when all authors are considered, productivity inequality is increasing over career ages while impact inequality, albeit larger, is mostly stable. Rising inequality in productivity is an effect of considering the full workforce and disappears for comparable authors, at least for cumulative counting.

\begin{figure*}[t!]
    \centering
    All authors\\
    Every Author assignment {} {} {} {} First Author assignment\\
    \includegraphics[scale=0.28]{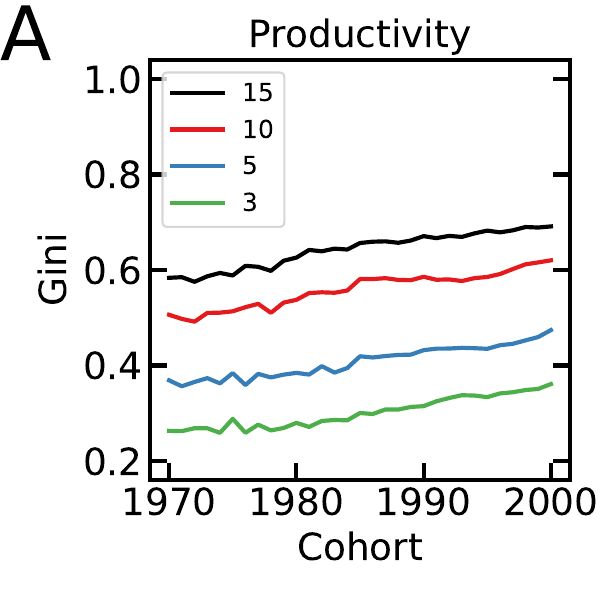}
    \includegraphics[scale=0.28]{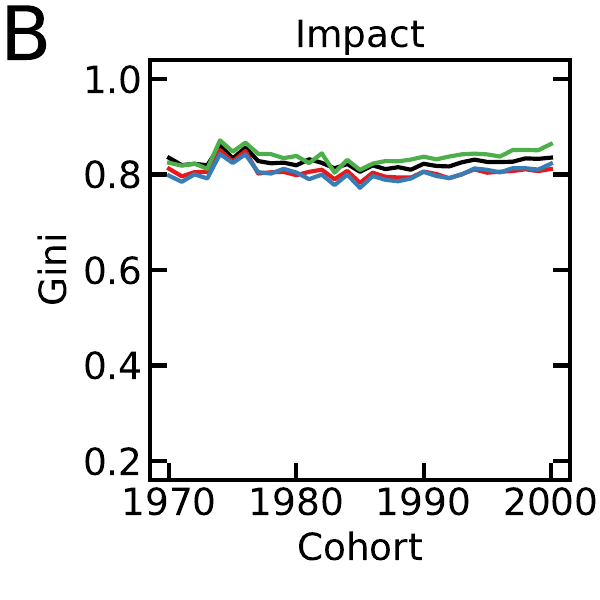}
    \includegraphics[scale=0.28]{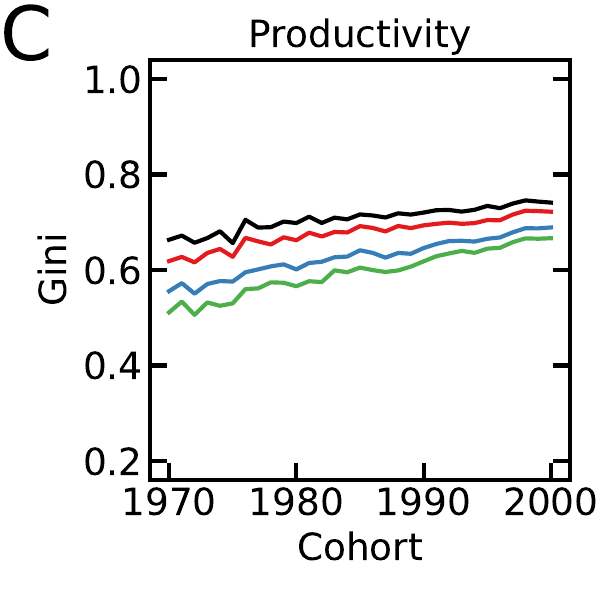}
    \includegraphics[scale=0.28]{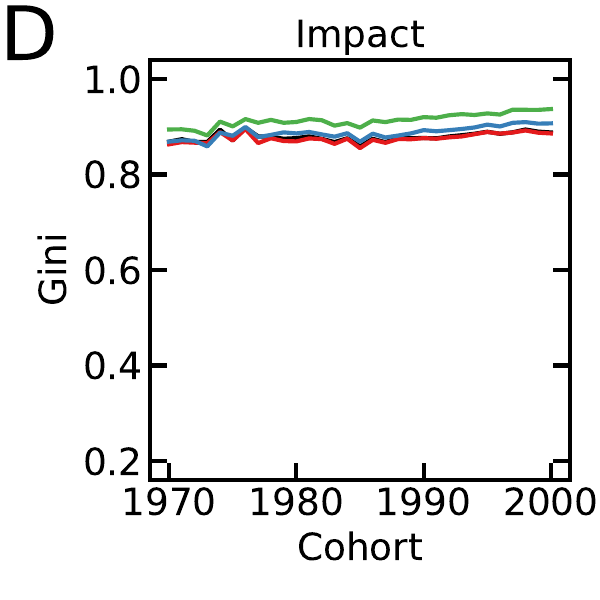}\\
    Dropouts removed\\
    Every Author assignment {} {} {} {} First Author assignment\\
    \includegraphics[scale=0.28]{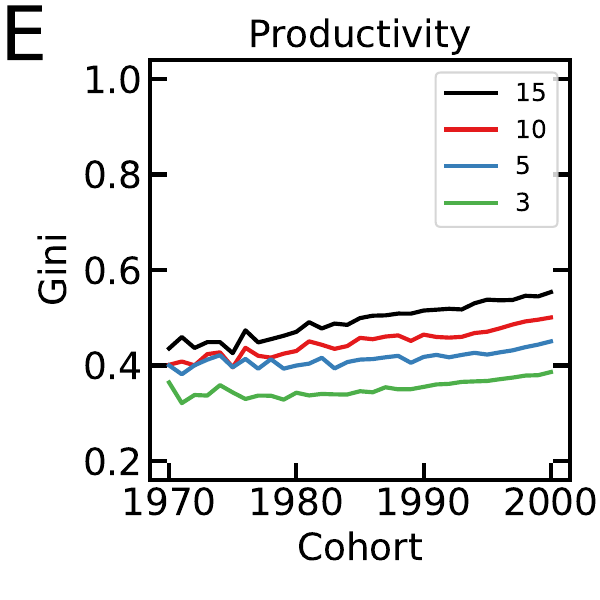}
    \includegraphics[scale=0.28]{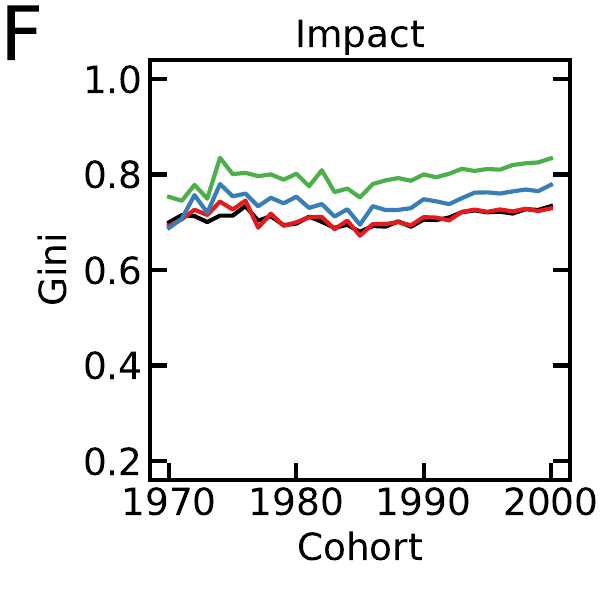}
    \includegraphics[scale=0.28]{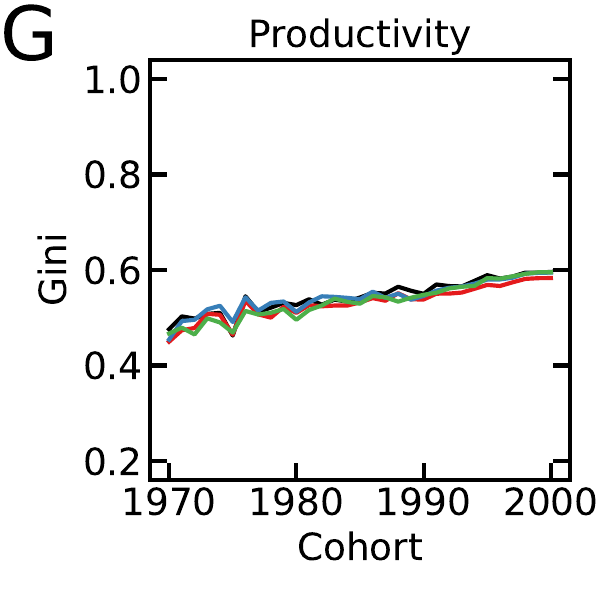}
    \includegraphics[scale=0.28]{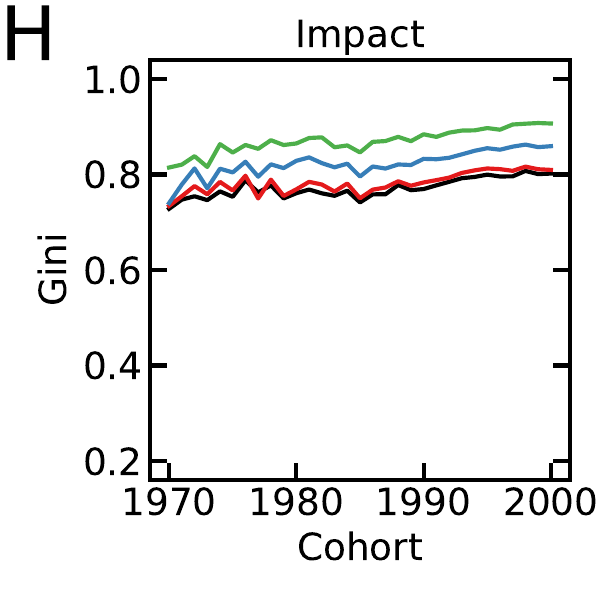}
    \caption{Individual inequality in productivity and impact as a function of cohorts, depicted for career ages 3, 5, 10, and 15.
    We count publications and citations cumulatively ($P(t)$ and $C(t)$, defined in ``Materials and methods: Individual inequality'').
    (First two columns) Assigning publications to all authors. (Last two columns) Assigning publications only to first authors.
    (Second row) Authors are filtered that have not published for ten consecutive years (most likely left academia).
    }
    \label{fig:gini-cum-across-cohorts}
\end{figure*}

Now turning to the historical analysis of cohorts, we address \citeauthor{zuckerman_age_1972}'s (\citeyear{zuckerman_age_1972}) hypothesis that recruiting more people into science will lead to larger differences between the most and the least talented.
Our results do not entirely support this hypothesis: we do not see a remarkable increase in inequality over cohorts for impact, but we observe an upward trend for productivity (figures \ref{fig:gini-cum-across-cohorts}A and B).
Removing dropouts reduces inequality levels but no longer softens the increasing trend for productivity (\ref{fig:gini-cum-across-cohorts}E). Also considering first authorships preserves all trends, this time also for productivity. The increase in productivity inequality does not vanish when only comparable authors are considered (\ref{fig:gini-cum-across-cohorts}G).
The results are not as evident for window counting of publications and citations because the Gini coefficients are much closer to 1 (appendix \ref{appendix}, figure \ref{fig:gini-win-across-cohorts}).
In sum, though the field has grown exponentially, similar levels of impact inequality can be observed for authors that started their careers in 1970 and 2000. When it comes to productivity, however, inequality appears to be increasing over time in parallel to the field's transition from an individual- to team-based science. 

\subsection{Gender inequality over careers and cohorts}
\label{sec_horizontal_time}

\begin{figure*}[t!]
    \centering
    All authors\\
    Every Author assignment {} {} {} {} First Author assignment\\
    \includegraphics[scale=0.28]{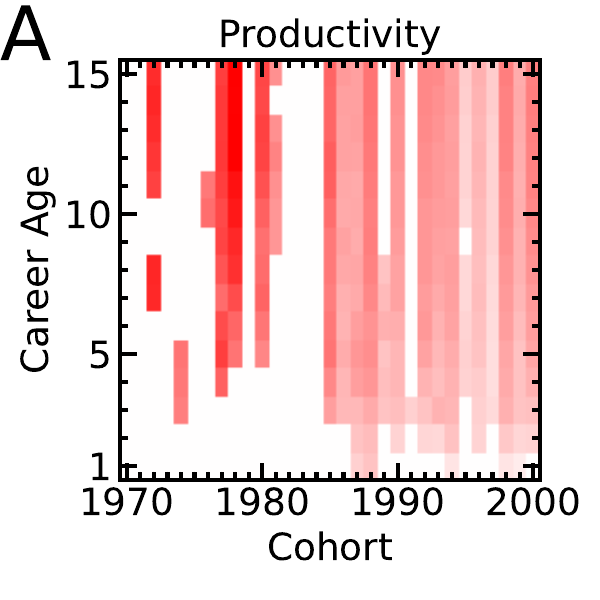}
    \includegraphics[scale=0.28]{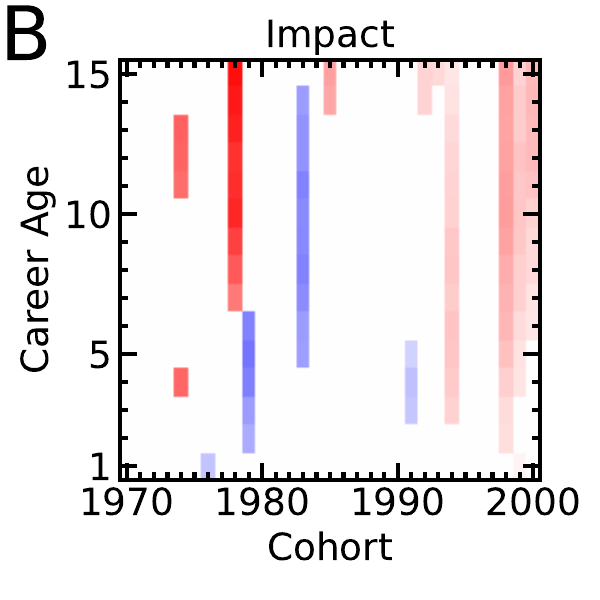}
    \includegraphics[scale=0.28]{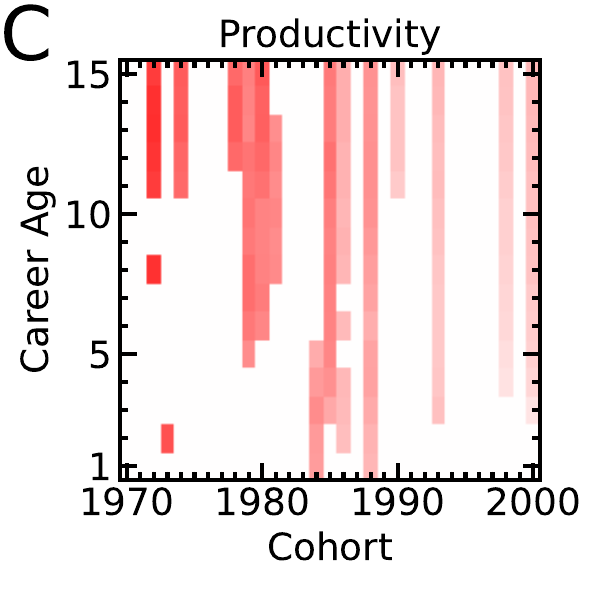}
    \includegraphics[scale=0.28]{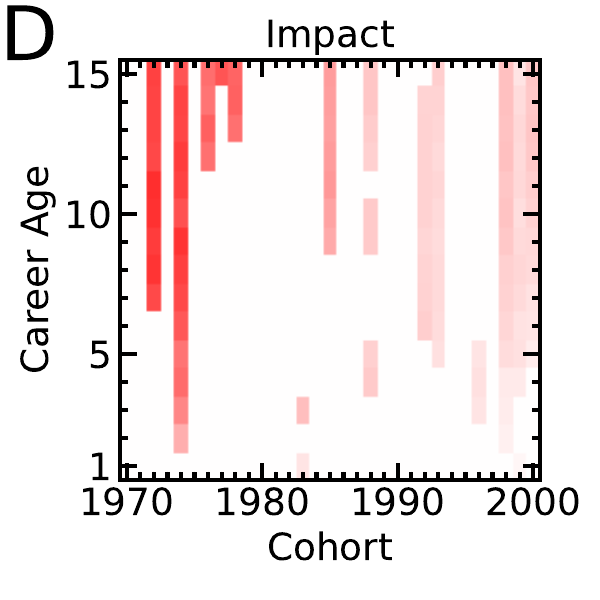}
    \includegraphics[scale=0.28]{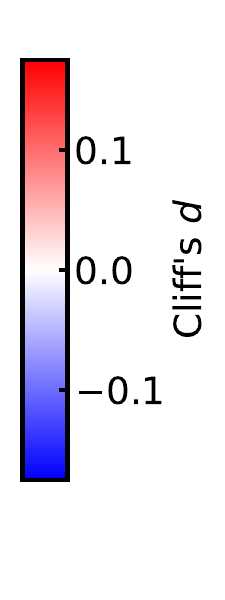}\\
    Dropouts removed\\
    Every Author assignment {} {} {} {} First Author assignment\\
    \includegraphics[scale=0.28]{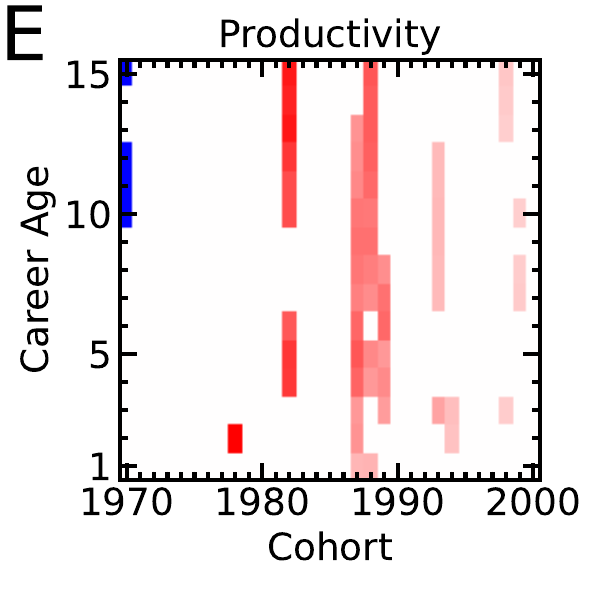}
    \includegraphics[scale=0.28]{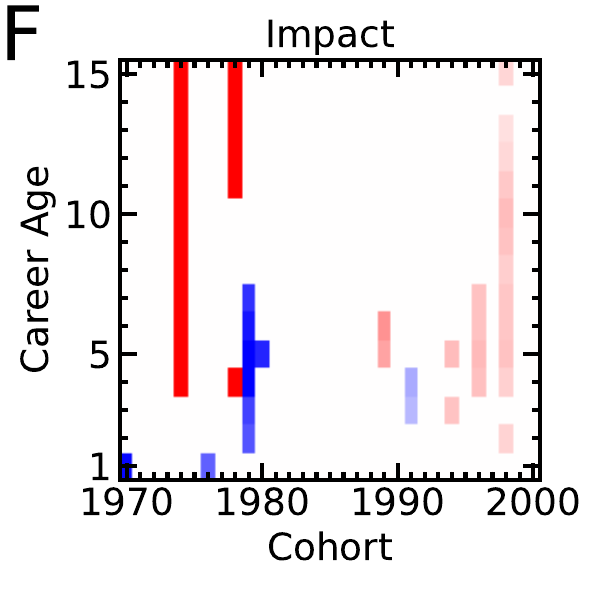}
    \includegraphics[scale=0.28]{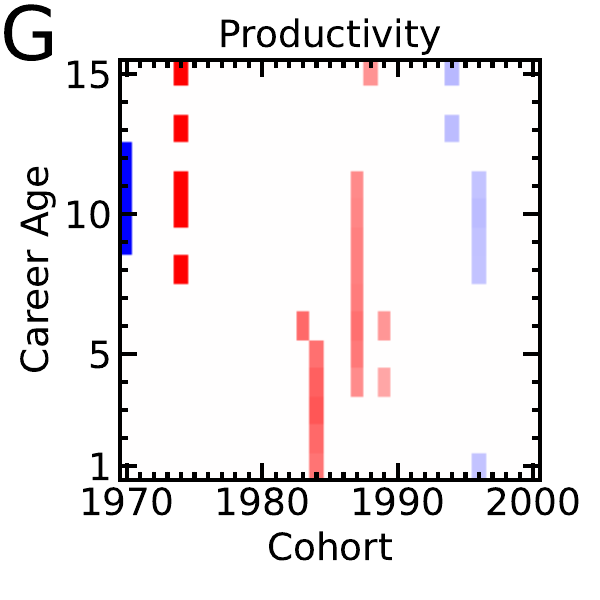}
    \includegraphics[scale=0.28]{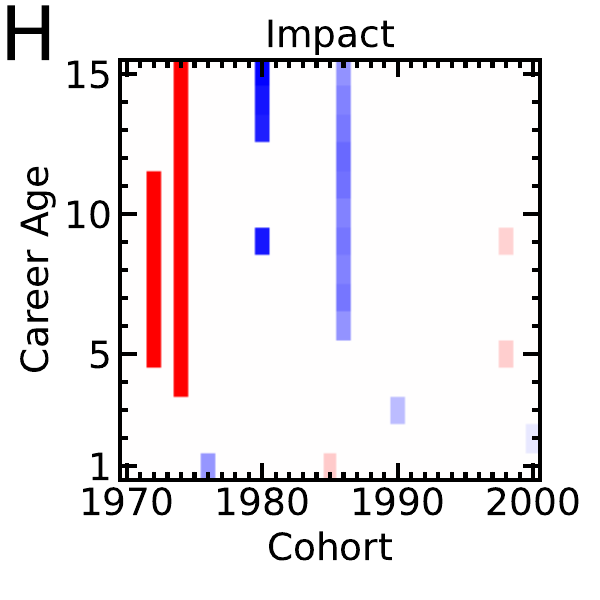}
    \includegraphics[scale=0.28]{figs/fig4_horiz_ineq_colorbar.pdf}
    \caption{Gender inequality for productivity and impact as a function of cohort and career ages.
    We compare the cumulative publications distribution $P_{\mathrm{gender}}(t)$ and cumulative citations distribution $C_{\mathrm{gender}}(t)$ of male and female scientists in the same cohort at the same career age $t$ and test differences between these distributions. Color marks the effect size (Cliff's $d$). Positive values (red) indicate that men dominate women, while negative values (blue) reveal that women dominate men. Effects are only shown if they are significant ($p\leq0.05$) according to a Mann–Whitney $U$ test. Details in ``Materials and methods: Gender inequality.'' Publications are assigned to all authors (A, B) or first authors only (C, D).
    In general, effects decrease with cohort and increase with career age.
    }
    \label{fig_horiz_ineq}
\end{figure*}

Increasing individual inequality in science is not necessarily problematic if the evaluation is based solely on merit rather than on functionally irrelevant factors such as gender, race, nationality, age, or class. Due to its societal importance, we focus on gender inequality.
Figure \ref{fig_horiz_ineq} shows a systematic comparison of the cumulative productivity and impact distributions of male and female computer scientists with the same filters applied as in the previous figure. Positive values (red) indicate that the distribution of men is dominant, that is, men are more productive or their work has a higher scientific impact. Negative values (blue) reveal that the distribution of women is dominant.

There is a general pattern for gender inequality in productivity (figure \ref{fig_horiz_ineq}A): it seems to accumulate and is more prevalent in the later career stages. 
If there are differences in productivity it is always men publishing more. This gender productivity gap exists in almost all cohorts.
For gender inequality in impact, the picture is less clear (\ref{fig_horiz_ineq}B). Female and male dominance both exist sporadically in cohorts. In four cohorts, women are statistically more likely to have more citations than men, for the 1982 cohort even for ten consecutive career years. There is no cohort in which gender inequality shifts signs, which means, it is always one cohort's gender that is dominant.
In total, gender inequality is more pronounced for productivity than for impact. For cumulative numbers of publications, $55\%$ of 465 cohort-age pair differences are statistically significant; for cumulative numbers of citations, $19\%$ are significant. That means, the productivity gap does not automatically translate into an impact gap. However, whenever there is an impact gap, it can be explained by a productivity gap: significant differences in citation are strongly correlated with differences in publications ($r=0.91, p\leq0.001$). That means, as \citet{azoulay_self_citation_2020} found, the productivity gap is the puzzle to solve.

When we limit authors to first authors, we get a step closer to solving this puzzle (figure \ref{fig_horiz_ineq}C): the magnitude of the gender gap becomes smaller (only $28\%$ of cohort-career year pairs are significantly different). This is particularly the case for the more recent cohorts.
The observation that larger differences in productivity between male and female scientists diminish when only first-author contributions are counted suggests that, as team sizes increased in computer science, male scientists boosted their productivity more via collaborations than female scientists. 
Applying the first author filter makes the impact gap a purely male phenomenon but also a phenomenon of the 70s (\ref{fig_horiz_ineq}D). 

Accounting for dropouts removes any pattern (\ref{fig_horiz_ineq}E to H). Neither does gender inequality increase with career ages nor does it persist on the historical scale or single out any gender. Any significant inequality is likely just noise.
In sum, gender inequality exists. While it appears to be diminishing on the timescale of cohorts, it is more persistent on the career timescale. Importantly, however, gender inequality practically disappears in recent cohorts when authors with comparable careers are studied.

\subsection{The role of the Matthew Effect}
\label{sec_me}

In the \emph{explanatory} modeling step that now follows, we inquire if reproductive feedback operates in the field as an underlying mechanism and to what extent it can generate the patterns of individual inequality we observe.
The ME states that present achievement (productivity or impact) depends on past achievement and that resulting advantages can accumulate over time.
Our guiding theory describes this feedback process as a vortex, an autocatalytic mechanism that fuels itself \citep{padgett_emergence_2012}. 
When the ME is fully operational -- formally, when it is linear -- it generates power law distributions that signal the absence of a characteristic scale \citep{albert_statistical_2002}.
In our case of computer science, productivity and impact distributions are broad but not pure power laws. Distributions for individual cohorts are much like the (truncated power law and stretched exponential) distributions that we measure when all cohorts are lumped together (figure \ref{fig_dataset_descriptives}C). These deviations can result from a damped (sublinear) ME and other mechanisms and factors that interact with the ME but also from sampling and finite-size effects intrinsic to the DBLP database.


\begin{figure*}[t!]
    \centering
    \includegraphics[scale=0.28]{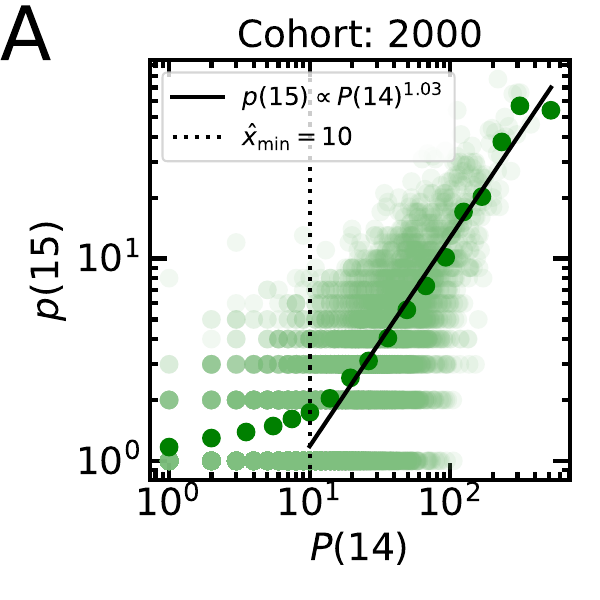}
    \includegraphics[scale=0.28]{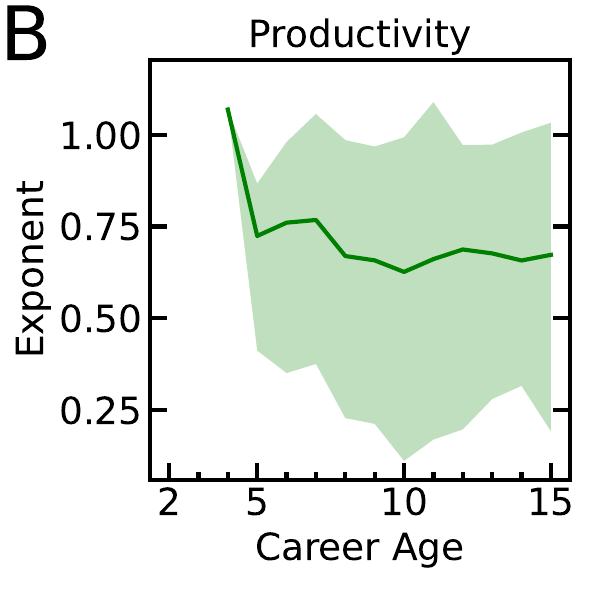}
    \includegraphics[scale=0.28]{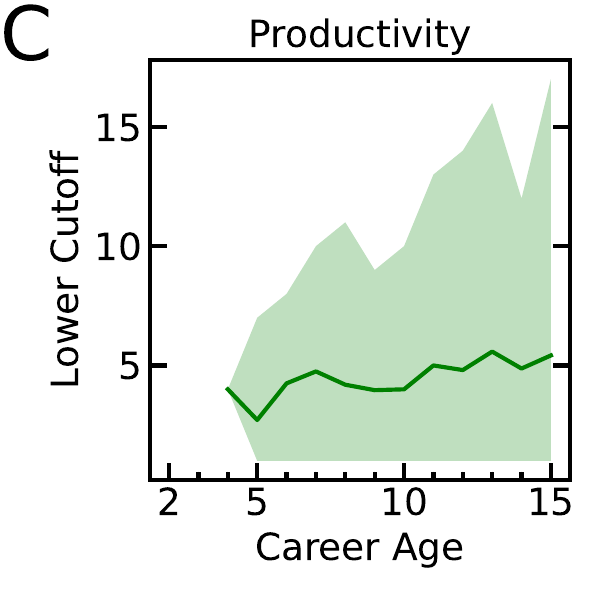}
    \includegraphics[scale=0.28]{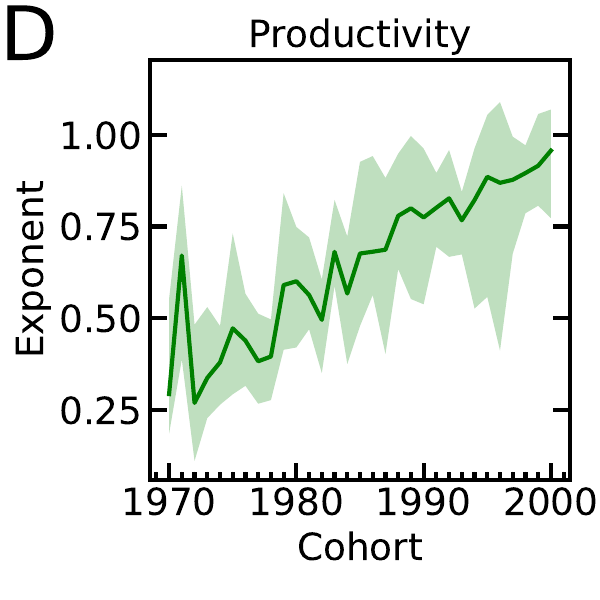}
    \includegraphics[scale=0.28]{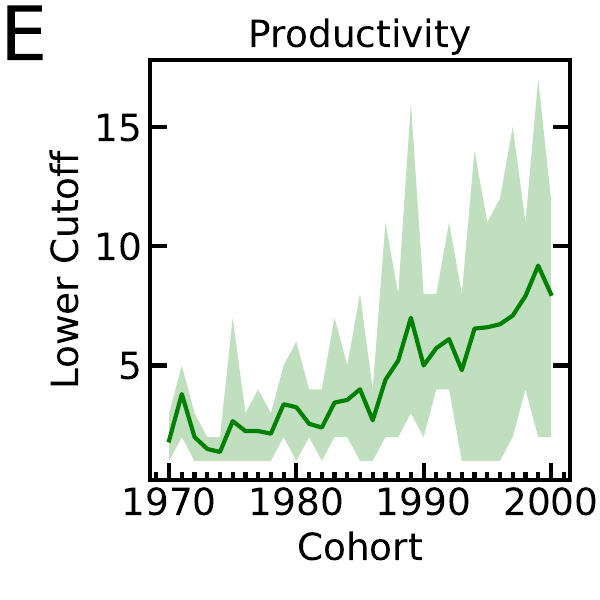} \\
    \includegraphics[scale=0.28]{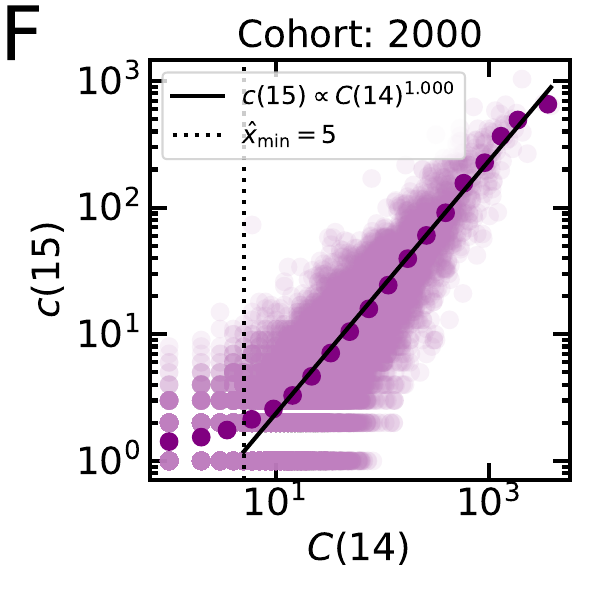}
    \includegraphics[scale=0.28]{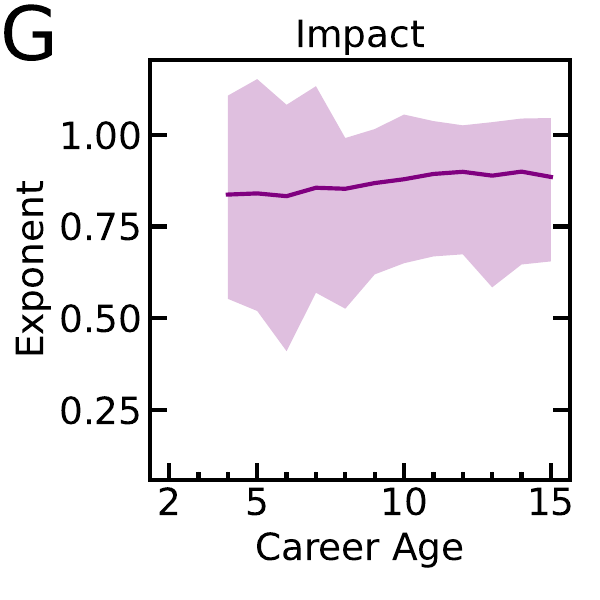}
    \includegraphics[scale=0.28]{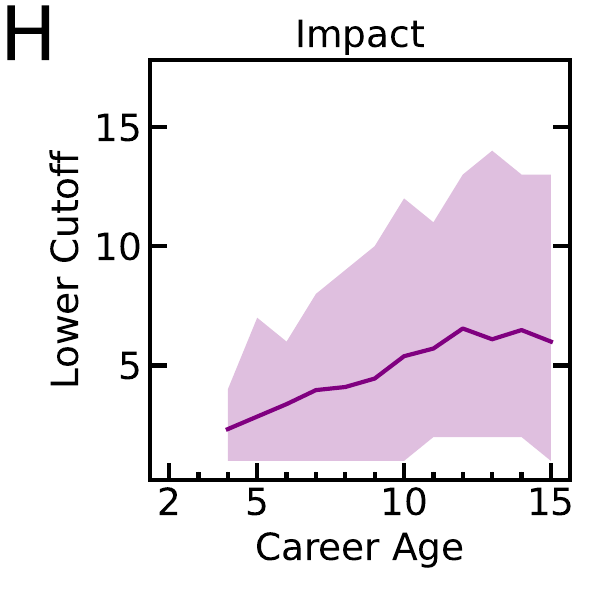}
    \includegraphics[scale=0.28]{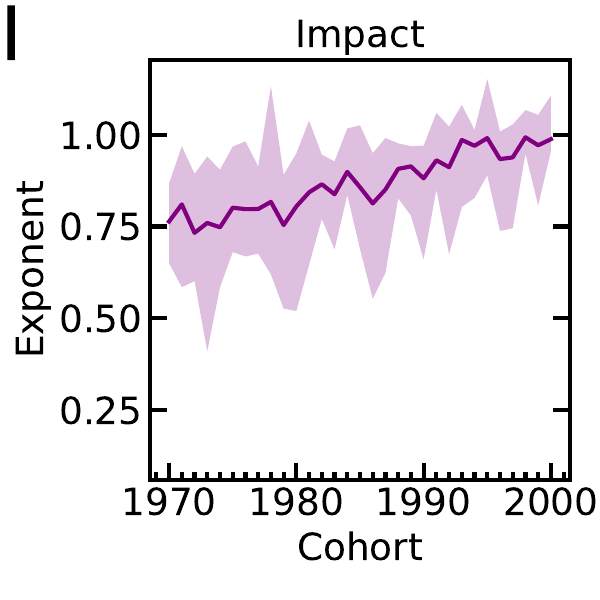}
    \includegraphics[scale=0.28]{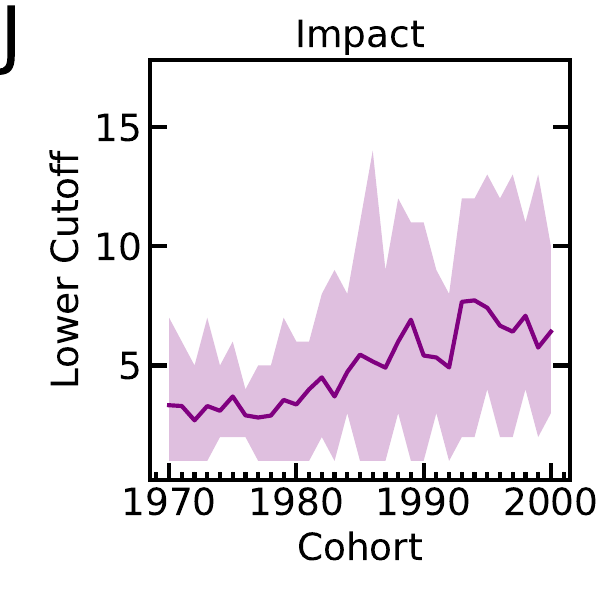} \\
    \caption{Matthew Effect.
    (First column: A, F) Measurement of the strength of a cohort's reproductive feedback as the exponent that relates an author's number of papers produced, or citations received, in a career age (y-axis) to the respective cumulative numbers in the previous career age (x-axis), shown for the 2000 cohort and the last career age. Exponents show as slopes of the continuous lines on the log-log plot. Dotted lines indicate that feedback fully unfolds only above a lower cutoff.
    (B, G) For an average cohort, potential individual advantages from feedback are constant along the career path, for both productivity and impact.
    (C, H) For an average cohort, the number of citations required to take advantage of feedback increases along the career path.
    (D, I) For an average career age, potential individual advantages from feedback increase historically, but more so for productivity.
    (E, J) For an average career age, the numbers of citations and publications required to take advantage of feedback increase historically.
    (All columns but the first) Shaded areas are bounded by minima and maxima, lines show means.}
    \label{fig_me}
\end{figure*}

We quantify the strength of the reproductive feedback of the field that a cohort experiences in a career age by regressing the number of publications or citations in a career age on the corresponding cumulative number in the previous career age (details in ``Materials and methods: Reproductive feedback'').
We interpret two parameters. The \textit{exponent} of the scaling relationship quantifies the strength of reproductive feedback. An exponent that is larger than zero over time is indicative of a cumulative advantage.
The \textit{lower cutoff} states at and above which number of publications or citations the advantage accruing from past selection unfolds. It resembles the boundary to the basin of attraction of the feedback dynamics: once an author crosses it, she or he gets attracted by the reproductive vortex and advantages can accumulate. 
Examples of the fitting procedure are depicted in figures \ref{fig_me}A and F. They show that scaling relationships are plausible fits to the data.

Our results show that the ME is a plausible explanation for productivity and impact inequality since all exponents are larger than zero. For an average cohort, the strength of the ME is stable over an author's career, allowing for a constant cumulative advantage. This holds for both productivity and impact as there are no discernible trends in figures \ref{fig_me}B and G. To enter the productivity basin of attraction (i.e., to reap benefits), an author must produce a certain number of publications that is constant over career ages (non-discernible trend in figure \ref{fig_me}C). However, getting one's publications cited becomes increasingly difficult as careers progress since the lower cutoff increases with career age (figure \ref{fig_me}H). In other words, regarding productivity, it is equally possible for an early- and late-career author to benefit from autocatalytic feedback, but regarding impact, moving early is advantageous.

While the strength of the ME is stable over a computer scientist's career, it does increase at the historical timescale of cohorts. Nowadays, the ME is strong for both impact and productivity (the exponents in figures \ref{fig_me}D and I are $\approx 1$ for the 2000 cohort start year). However, whereas the 1970 cohort already experienced a strong effect from past citations (exponent $\approx0.8$), the effect of the past number of publications started weak ($\approx0.3$). In other words, while getting cited has long been endowed with a strong reinforcement effect, increasing returns for productivity became prominent only recently. At the same time, the lower cutoff for reinforcement to set in has been growing historically, particularly so for productivity (figure \ref{fig_me}J). As the field grew and transitioned towards team-based science, this is likely the result of how limited resources get distributed among an increasing number of scholars. For the authorship practice, this mechanism has a name: ``publish or perish'' \citep{garfield_what_1996}.


There is a correspondence between the ME and the inequalities described in figures \ref{fig:gini-cum} and \ref{fig:gini-cum-across-cohorts}: The ME is persistently stronger for impact than for productivity and thus generates individual inequalities that are persistently higher, both over careers and cohorts.\footnote{Since we have measured the ME cumulatively we do not discuss a correspondence with the window-counted inequalities.} Looking at cohorts for average career ages, a large cumulative advantage for productivity corresponds to a modest increase in inequality while a modest cumulative advantage for impact corresponds to stable inequality. For careers, there is no meaningful correspondence. We suggest that this is because, for a certain year, the ME is always computed for authors that have been active in that year, that is, that they have either published or got cited. This means that, in figures \ref{fig_me}B, C, G, and H, the values for small career ages result from all authors while the values for large career ages tend to result from authors with dropouts removed.
In sum, the ME increases, strongly so for productivity where it resembles the imperative to "publish or perish." Reproductive feedback is stronger and creates larger inequality for impact than for productivity but, alone, is not capable of explaining the individual inequality patterns we observe.  

\subsection{The role of gender}

\begin{table}[b!]
    \caption{Independent variables used in the prediction models. The variables characterize authors in their early career ages $[1,t_\mathrm{e}]$, with the end of the early career chosen to be $t_\mathrm{e}=3$. The variables are used to predict author dropout and success. Details are given in the section ``Materials and methods''.}
    \label{tab:variables}
    \centering
    \resizebox{\textwidth}{!}{%
    \begin{tabular}{ll}
    \hline
    \textbf{Variable} & \textbf{Description} \\ \hline
    \multicolumn{2}{l}{\textbf{Baseline}} \\
    Cohort & Year in which cohort members started publishing \\ \hline
    \multicolumn{2}{l}{\textbf{Gender}} \\
    Male  & Dummy \\
    Female  & Dummy \\
    Undetected  & Dummy \\ \hline
    \multicolumn{2}{l}{\textbf{Early Achievement}} \\
    Productivity & Cumulative number of publications authored in the early career \\
    Productivity (1st author) & Cumulative number of publications authored in the early career as a first author \\
    Impact & Cumulative number of citations received in the early career \\ 
    Top source &  Smallest $h5$-index-based quartile rank of all journals and conference proceedings an author has published in the early career \\ \hline
    \multicolumn{2}{l}{\textbf{Social Support}} \\
    Collaboration network & Number of distinct co-authors in the early career \\
    Team size & Median number of authors of all publications produced in the early
career \\
    Senior support & Largest $h$-index of all co-authors in the early career \\ \hline
    \hline
    \end{tabular}
    }
\end{table}

Following the leads from the descriptive and explanatory analyses, we proceed with the final analytical step: out-of-sample \textit{predictions} of dropout and future success. We study the effect of the cohort, gender, and variables from two classes of constructs. First, our theory of careers states that inequality results from differences in the early career that are then amplified by the ME which we have found to operate. We aim to gain more insights into how the ME operates by using a set of variables on the early-career achievements of authors. Second, we study inequality as the field becomes a more team-based science. We have found that it is historically increasingly important to produce publications to benefit from the ME. Hence, we use a set of variables on social support which, we hypothesize, makes it easier to write papers. The constructs are fully described and operationalized in the section ``Materials and methods: Independent variables'' and summarized in table \ref{tab:variables}. We use regression models with standardized variables so that the coefficient size indicates the relative importance of the predictor. The models also employ variable selection. When new variables are added to the model, this can result in decreasing effect sizes for previously considered variables. If that happens, there is collinearity among the variables and the variable that predicts better has a larger coefficient.

\subsubsection{Predicting dropout}
\label{sec:sec_dropouts}
\begin{table}[b!]
    \caption{Dropout prediction.
    Each column corresponds to a separate logistic regression model that predicts whether (1) or not (0) an author dropped out of computer science (described in section ``Materials and methods: Prediction models''). Coefficients are reported as means (with standard deviations in brackets) from 10-fold cross-validation over 292,443 observations. Mean coefficients away from zero indicate that the effects are sizable, and low standard deviations indicate that the effects are robust. Goodness-of-fit measures (F1 and average precision) are also means across all folds. 
    }
    \label{tab:aggr_dropouts}
    \centering
    \small
    \setlength\tabcolsep{2pt}
    \renewcommand\arraystretch{1.1}
    \begin{tabular}{lrrrr}
    \hline
     & \textbf{Model 1} & \textbf{Model 2} & \textbf{Model 3} & \textbf{Model 4} \\ 
     & \textit{Baseline} & \textit{+ Gender} & \textit{\begin{tabular}[c]{@{}c@{}}+ Early\\ achievement\end{tabular}} & \textit{\begin{tabular}[c]{@{}c@{}}+ Social\\ support\end{tabular}} \\ 
     \hline
    Cohort             & 0.00(0.00) &   0.00(0.00) &        0.00(0.00) &     0.00(0.00) \\
Female             &            &   0.03(0.01) &        0.06(0.01) &     0.06(0.01) \\
Male               &            &  -0.06(0.01) &       -0.07(0.00) &    -0.08(0.00) \\
Undetected         &            &   0.03(0.00) &        0.02(0.00) &     0.02(0.00) \\
Productivity       &            &              &       -0.56(0.00) &    -0.54(0.00) \\
Productivity (1st) &            &              &       -0.26(0.00) &    -0.22(0.01) \\
Impact             &            &              &        0.01(0.00) &     0.01(0.00) \\
Top source         &            &              &       -0.22(0.01) &    -0.21(0.01) \\
Collaboration network &         &              &                   &    -0.08(0.00) \\
Senior support     &            &              &                   &    -0.01(0.00) \\
Median team size   &            &              &                   &     0.22(0.01) \\\hline
Intercept          &       0.00 &         0.30 &              0.00 &           0.00 \\
F1                 &       0.44 &         0.44 &              0.67 &           0.68 \\
Average precision  &       0.58 &         0.60 &              0.75 &           0.76 \\
\hline \hline

    \end{tabular}
\end{table}

Table \ref{tab:aggr_dropouts} shows the results of logistic regression models that use dropout as a binary dependent variable. The most important factor for predicting dropout is early career productivity. Scientists that publish much in the first three years of their careers, not necessarily as a first author, are less likely to drop out. This is not surprising, given that dropout is defined as the absence of publications for ten consecutive career years. 
Publishing in a top source early on is the second strongest predictor of not dropping out. Having a publication in a top journal or conference proceedings represents symbolic capital, a reputation signal for academic worthiness, that likely influences the career path. These differences in early-career achievements are then amplified by the ME, contributing to the individual inequalities we have diagnosed. Social support has effects that seem contradictory at first glance. On the one hand, co-authoring publications in large teams is positively correlated with dropout. On the other hand, having larger collaboration networks decreases the likelihood to drop out. This tells us that being one author among many is not automatically an achievement. It is the well-connected authors that stay in the field. Having early senior support (a co-author with a high $h$-index) has negligible influence on dropping out.
Similarly, dropout is not associated with the initial impact of early-career publications. Having many early citations does not make it more likely for an author to stay in computer science.
Women are more likely to drop out than an average computer scientist and adding early achievements and social support even increases this effect. Interestingly, the cohort has no effect. These dropout-related patterns are historically invariant. That means, large author teams did not guard against dropping out before and in the team era.

\subsubsection{Predicting success}

\begin{table}[b!]
    \caption{Success prediction.
    Each column corresponds to a separate linear regression model that aims to predict $C_i^+(15)$, the increase in citations an author gains after the early career of the first three career years (described in section ``Materials and methods: Prediction models''). Coefficients are reported as means (with standard deviations in brackets) from 10-fold cross-validation over 292,443 observations. Mean coefficients away from zero indicate that the effects are sizable, and low standard deviations indicate that the effects are robust. Goodness-of-fit measures (mean squared error and adjusted $R^2$) are also means across all folds. In the first four models, there are 292,443 observations. In model 5 dropouts are removed which causes the number of observations to drop to  119,113.
    }
    \label{tab:aggr_citations}
    \centering
    \small
    \setlength\tabcolsep{2pt}
    \renewcommand\arraystretch{1.1}
    \begin{tabular}{lrrrrr}
    \hline
     & \textbf{Model 1} & \textbf{Model 2} & \textbf{Model 3} & \textbf{Model 4} & \textbf{Model 5} \\ 
     & \textit{Baseline} & \textit{+ Gender} & \textit{\begin{tabular}[c]{@{}c@{}}+ Early\\ achievement \end{tabular}} & \textit{\begin{tabular}[c]{@{}c@{}}+ Social\\ support\end{tabular}} & \textit{\begin{tabular}[c]{@{}c@{}}Dropouts\\ removed\end{tabular}} \\ 
     \hline
    Cohort             & 0.08(0.00) &   0.08(0.00) &        0.07(0.00) &     0.05(0.00) &     0.04(0.00) \\
Female             &            &  -0.07(0.11) &       -0.09(0.01) &    -0.10(0.01) &     -0.04(0.01) \\
Male               &            &   0.01(0.02) &        0.00(0.00) &     0.00(0.00) &     0.00(0.00) \\
Undetected         &            &   0.00(0.00) &        0.06(0.01) &     0.04(0.01) &     0.03(0.01) \\
Productivity       &            &              &        0.99(0.01) &     0.88(0.01) &     0.45(0.01) \\
Productivity (1st) &            &              &        0.44(0.01) &     0.45(0.01) &     0.33(0.01) \\
Impact             &            &              &        1.00(0.01) &     0.91(0.01) &     0.37(0.01) \\
Top source         &            &              &        0.19(0.01) &     0.00(0.00) &     0.05(0.00) \\
Collaboration network &            &              &                   &     0.05(0.01) &     0.02(0.01) \\
Senior support     &            &              &                   &     0.70(0.01) &     0.40(0.01) \\
Median team size   &            &              &                   &    -0.09(0.01) &     -0.05(0.01) \\\hline
Intercept          &    -158.94 &      -157.86 &           -131.24 &         -94.95 &     -77.45 \\
Mean squared error &      40.17 &        40.17 &             31.95 &          31.54 &     7.28 \\
Adjusted $R^2$     &       0.01 &         0.01 &              0.21 &           0.22 &     0.22 \\
\hline \hline

    \end{tabular}
\end{table}

Next, we study which factors predict whether success in terms of scientific impact increases, on average, after the first three career years (table \ref{tab:aggr_citations}). The dependent variable is the increase in citations until career age 15.
First of all, the cohort has a small positive effect on success. This can be an effect of the exponential growth of the field: As more publications are produced and reference lists become longer, more citations are made and accumulated \citep{Pan2018}. 
Early-career productivity is a requirement for success, and its effect is on par with that of early-career impact. This confirms an observation made in the literature, namely that total success is well predictable from early success \citep{mazloumian_predicting_2012, penner_predictability_2013, wang_quantifying_2013}. Since the ME is path-dependent this is further evidence for cumulative advantage as an underlying mechanism.
In contrast to dropout, early senior support is an important factor for success. But similar to dropout prediction, publishing in large teams exhibits a negative effect on citation success, and the size of the early collaboration network has a weakly positive effect. Publishing in a top source is associated with success but gives way to senior support when added to the model.

Finally, being female is a negative predictor for an increase in citations, and the effect increases when we account for early achievements and social support. This adds to works that have also found gender inequality in impact \citep{Cole1984, lincoln_matilda_2012, Cassidy2013}.
To check if this effect can again be explained by the gender difference in career persistence, we construct a last model that predicts success with dropouts removed. The result is reported in the last column of table \ref{tab:aggr_citations}. Removing dropouts makes the absolute values of all effects smaller (except for publishing in a top source). This means that persistent authors are more homogeneous in their characteristics. The markedly smaller error also shows that their success is easier to predict.
However, the gender effect does not fully go away: After removing dropouts, women are still slightly less successful at career age 15 than an average computer scientist.

\section{Discussion}
\subsection{Summary and conclusion}

We studied individual and gender inequalities in computer science, their changes over author careers as well as over the field's transformation from a sole-scholar to team-based science, and their origins. We found that individual inequality in productivity increases during the careers of an average cohort but, contrary to what has been previously suggested \citep{Allison1982}, it does not translate to an increase in impact inequality. The increase in productivity inequality can be a result of comparing scholars with different persistence (all authors vs. only those that kept producing papers) and status (all authors vs. only first authors of a paper), but this explanation is not robust to changing the counting method.
The inequality patterns of cohorts from 1970 to 2000 are different from those of chemistry cohorts in the 1960s and 1970s.
Since computer science exhibits an exponential influx of personnel, we have also checked if it leads to an increase in individual inequality on the historical time scale of cohorts \citep{zuckerman_age_1972}. We found such an effect but only a small one for publications. The inequality patterns for impact are largely the same since the 70s.

Regarding gender inequality, we found that men produce more publications than women, particularly towards the end of their careers, though this phenomenon was more pronounced in the past. This gender gap, known as the ``productivity puzzle'' \citep{reskin_scientific_1979, Cole1984, cole_theory_1991, Long1992}, disappears once we remove dropouts and focus on first authors only.
Regarding reports of gender inequality in impact \citep{Cole1984, lincoln_matilda_2012, Cassidy2013}, we found that men having more publications does not automatically entail having more citations, but more citations find their explanation in more publications.

To understand individual inequality, we quantified the Matthew Effect (ME). We found that it is stable over an author's career regarding productivity and impact. While the number of publications above which nonlinear benefits accrue is also stable as the career progresses, this becomes increasingly difficult regarding the number of citations, indicating an early-citation advantage. The ME for publications and citations has increased historically to similar levels, but the climb has been much steeper for productivity, indicating the rise of the imperative to "publish or perish."

Using regression models to inquire about the importance of early career achievements and social support in shaping total-career outcomes, we found that early productivity is the best predictor for not dropping out of computer science and for impact-based success. While staying in the field is a condition for success, success and having successful co-authors are not conditions for staying in the field.
As our result for the ME suggested, success turned out to be correlated with early-citation success. Publishing in top journals and conference proceedings in the early career is predictive of staying in the field as well as success.
Authors with a large social support network are more likely to stay and be successful, stressing the importance of social capital, but authors that are part of large co-author collectives are more likely to drop out or remain unsuccessful, potentially because more of the same in terms of team structure hinders creativity \citep{uzzi_collaboration_2005, guimera_team_2005}. This supports the argument that the transition to team science is a reason why we observe a historical increase in individual inequality for productivity.

Finally, we found that women are not only more likely to drop out of the field but also somewhat less successful after 15 years than an average computer scientist. Both effects cannot be explained with less impressive early-career achievements or lower social support. 
Whereas gender inequalities could be explained when only comparable authors were analyzed, this also does not explain the impact gender effect: it is softened but does not go away when dropouts are removed.
The gender effects can potentially be explained by differences in network structure. We have found in previous work that, on average, "female" collaboration networks are smaller and more cohesive than the networks of their male counterparts \citep{Jadidi2017}. If this type of embedding is a disadvantage, the latter would accumulate due to the ME. For example, if men manage to inflate their publication counts more than women due to having different social capital \citep{Way2016}, this can explain our finding that the productivity gap between women and men is smaller when only publications authored as first authors are counted.

In conclusion, we have contributed to the reconstruction of the chain of events that results in individual and gender inequality in computer science. Teams become more important. These help scholars increase the number of publications which serves their career. But being part of a team is not enough, only some authors manage to reap benefits from team science and the cumulative advantage of early-career achievements. Most scholars drop out of the field, especially women due to reasons we cannot measure. Differences in dropout entail differences in individual and gender inequality. Taken together, in the 45 years we have studied, computer science has increasingly become a competitive field. At the end of a career that works like a tournament, senior female computer scientists even fall behind in terms of citation impact to some extent.

These conclusions in the context of computer science likely carry over to other male-dominated fields that have experienced growth and transformation from individual-based to project- and team-based science, such as statistics, applied mathematics, and engineering. Nevertheless, our findings should be replicated for other academic fields to establish the extent to which the trends in individual and gender inequality we detected are sensitive to the rates of growth, dropout, or entry of women.  

Our findings are relevant for science policy measures that aim at more gender equality.
The way the field operates on autocatalytic feedback, natural differences among women and men like motherhood, but also small behavioral differences in the ways women and men embed into collaboration networks, can have large career consequences. Regarding behavioral differences, our results suggest that mentorship programs with the goal ofOn "broadening and institutionalizing women's support networks" \citep[][p.175]{abbate_recoding_2017} are promising. It is not unexpected that small policy interventions can have large consequences, too.

\subsection{Methodological considerations}

We close our paper with a few methodological considerations. 
The first set of considerations relates to data. Inquiries into inequality and the ME date back to the 70s when it was only possible to study small cohorts. Much of this research was done using one of two carefully constructed bibliographic chemistry cohort datasets \citep{Allison1982}. On the contrary, we use a large-scale dataset on the complete trails of computer science. The use of bibliographic traces has allowed us to reconstruct and study scholarly careers in historical comparison. While formal communication just represents the observable aspect of academic careers, it is undeniably an important part since academic careers are subject to collective field dynamics that work on what is observable.

The ability to model processes with behavioral data comes at the cost of a reduced ability to model individual perception. Gender is an example. While we maintain that our inferred gender variable is a true gender variable because authors are free to choose which name they put on a paper, the variable does not allow us to differentiate between the various types of socially constructed gender. Hence, we can only contribute insights into the social construction of binary gender. In particular, we show indirectly how a structural mechanism that accumulates achievements -- the ME -- contributes to generating gender disparities, even after we account for early achievements and social support. Augmenting behavior with data on cognitive states (e.g., whether computer scientists dropped out on free terms, because of structural constraints or even discrimination) would allow for deeper insights into the origins of inequality.

Limitations to gender disambiguation from names forced us to remove most Asian authors from the analysis. However, the proportion of Asian scholars in computer science has been increasing since the 70s as a result of the passing of the 1965 Immigration Act in the US and the increasing internationalization of science. We are thus missing a larger proportion of one type of authors. Yet, the DBLP data has the converse problem too because its coverage of publications increases over time. We recognize that both of these biases might be influencing the historical trends we observe, and hence, our conclusions should be interpreted with caution.

Operationalizing scientific impact via the number of citations is straightforward because it very well captures that impact is a collective phenomenon. On the other hand, citation scores are not unobtrusive measures anymore. Citations have become a currency in science, scholars try to improve their scores, and the databases we use for research are also used to compute a scholar's market value. This adds an exploitative angle to the ME that cannot be disentangled from its systemic effects \citep{xie_undemocracy_2014, clauset_datadriven_2017}. And yet, strategic human behavior is the result of structural constraints as well as incentives and, hence, just as much part of the problem of inequality. 

The second set of considerations relates to methods. We followed the integrative modeling approach \citep{hofman_integrating_2021} and found the combination of descriptive, explanatory, and predictive modeling insightful. With large-scale behavioral data, exploratory description is necessary because existing knowledge may not translate into meaningful research questions or hypotheses for testing: past small-scale studies may not have captured new phenomena, while new large-scale studies may not generalize due to preprocessing decisions and design choices. Still, we let the literature guide our modeling: we distilled a theory of careers from a broad and multidisciplinary set of studies that gives the ME the central role of an inequality-creating mechanism. 
Multivariate regression models with variable selection then served to shed new light on findings from the first two modeling steps by focusing on early-career factors. This fleshed out the inequality-creating mechanism but also uncovered the gender impact effect even though gender inequality in impact had not been diagnosed.
The strong message is that the choice of methods matters and that a systematic mix of methods can produce more robust as well as surprising results.

\section{Materials and methods}


\paragraph{Data:}
We use DBLP \citep{DBLP:journals/pvldb/Ley09, dblp}, a comprehensive collection of computer science publications from major and minor journals and conference proceedings. From this dump, we remove \textit{arXiv} preprints. The coverage of DBLP ranges from $55\%$ in the 80s to over $85\%$ in 2011 \citep{Way2016}. Our dataset consists of 2.5 million 
publications from 1970 to 2014 that are authored by 1.4 million
authors. Of those, 292.443 started their career between 1970 and 2000.
We have added citations among publications by combining DBLP with the AMiner dataset \citep{aminer_paper, aminer_data} via publication titles and years.
There are 7.9 million 
citations among publications. 
Author names in DBLP are disambiguated \citep{DBLP:series/lnsn/Reitz013}.

To infer the gender of authors, we have used a method that combines the results of name-based (genderize.io) and image-based (Face++) gender detection services. The accuracy of this method is above $90\%$ for most nationalities. Since the accuracy is very low for Chinese and Korean names, we label their gender as unknown to reduce noise in our analysis \citep{Karimi:2016}.
Since authors are free to choose the name under which they publish, the inferred variable is a true, socially constructed gender attribute.

\paragraph{Cohorts and career ages:}
Our main units of analysis are cohorts of computer scientists from 1970 to 2000. We consider a career to begin with an author's first publication in the database. Since DBLP covers publication years back to 1960, this ensures that authors of the earliest cohort have been at least absent for ten years. Imbalances in coverage over publication years cause earlier cohorts to be less homogeneous as we tend to miss more first publications.
Given start years, we follow cohort members over career ages $t\in[1,15]$.

\paragraph{Publication and citation counts:}
Our unit of observation is the individual author $i$ in a cohort. For each author and career age, we measure the number of publications $p_i(t)$ authored in a career age, the cumulative number of publications $P_i(t)$ authored until and in a career age, the number of citations $c_i(t)$ received by $P_i(t)$ in a career age, and the cumulative number of citations $C_i(t)$ received by $P_i(t)$ until and in a career age.
Citations are always counted coming from the whole field of computer science, not just from the same cohort.

\paragraph{Individual inequality:}
To quantify individual inequality we use the Gini coefficient $G(t)$ of the publication and citation distributions of authors in the same cohort at the same career age:
\begin{equation}
    G(t)=\frac{\sum_{i=1}^{n} \sum_{j=1}^{n} |x_i(t)-x_j(t)|}{2n\sum_{i=1}^{n}x_i(t)}
\end{equation}
The numerator is the absolute difference of all pairs $(i,j)$ of authors in a cohort.
$x$ is a placeholder for publication or citation counts.
In figures \ref{fig:gini-win} and \ref{fig:gini-win-across-cohorts} in appendix \ref{appendix}, we use backward-looking 3-year windows, that is, to quantify inequality in productivity, 
$x(t)
=p_{\mathrm{3yr}}(t)
=\sum_{\tau=0}^{2}p(t-\tau)$, 
and, to quantify inequality in impact, 
$x(t)
=c_{\mathrm{3yr}}(t)
=\sum_{\upsilon=0}^{2}\sum_{\tau=0}^{\upsilon}c_{t-\upsilon}(t-\tau)$, 
where $t\geq3$ and the index $t-\upsilon$ of $c$ defines the career age for the publications of which citations are counted.
In figures \ref{fig:gini-cum} and \ref{fig:gini-cum-across-cohorts}, we use cumulative counting, that is, to quantify inequality in cumulative productivity, $x(t)=P(t)$, and, to quantify inequality in cumulative impact, $x(t)=C(t)$.
A Gini coefficient of zero expresses perfect equality, where all authors in one cohort have produced an equal number of papers or received an equal number of citations. A Gini of one indicates maximal inequality among authors.

\paragraph{Gender inequality:}
To quantify gender inequality, we look at the differences between the cumulative distributions of productivity $x(t)=P(t)$ and impact $x(t)=C(t)$ of male and female scientists in the same career age. 
For both $x(t)$, we rank all observations ascendingly (with adjusted ranks for ties) and perform the Mann–Whitney $U$ test,
\begin{equation}
U(t)=R_m(t)-\frac{n_m(t)(n_m(t)+1)}{2},
\end{equation}
where $R_m(t)$ is the sum of the ranks and $n_m(t)$ is the number of male scientists. The $U$ test allows us to assess the statistical significance of the difference between the distributions of male and female scientists \citep{mann_test_1947}.
To quantify the size of the difference, we compute Cliff's $d$,
\begin{equation}
d(t)=\frac{2U(t)}{n_m(t)n_f(t)}-1,
\end{equation}
where $n_f(t)$ is the number of female scientists \citep{cliff1993dominance}.
The value of $d$ ranges from $-1$ (when all observations for women are greater than those for men) to $1$ (when all observations for men are greater than those for women). For example, if $d=0.8$ for the cumulative publication distribution, a randomly picked man has an 80\% chance to have more publications than a randomly chosen woman. If $d=-0.8$ then a randomly picked woman has an 80\% chance to have more publications than a randomly picked man.

\paragraph{Reproductive feedback:}
We quantify the ME as the extent to which authors reproduce their productivity and impact over time via positive feedback.
For each cohort and career age, we measure to what extent scholars author new publications or receive new citations in a career age proportional to their productivity or impact in the previous career age. This relationship is quantified by the scaling law
\begin{equation}
    x(t)\propto x(t-1)^{\beta(t)}, x(t-1)\geq x_{\mathrm{min}}(t-1),
\end{equation}
where the exponent $\beta$ and the lower cutoff $x_{\mathrm{min}}$ are the model parameters. If the scaling law is a plausible fit and the estimated exponent $\hat{\beta}>0$, past productivity or impact is advantageous to, because correlated with, present productivity or impact. If this advantage accumulates over subsequent career ages, we speak of the ME that is then quantified by the sequence of $\hat{\beta}$s.
To quantify the ME in productivity we predict the number of publications $x(t)=p(t)$ by the cumulative number of publications $x(t-1)=P(t-1)$, and to quantify the ME in impact we predict the number of citations $x(t)=c(t)$ by the cumulative number of citations $x(t-1)=C(t-1)$. Predicting by the number of publications $p(t-1)$ and citations $c(t-1)$ yields less variance in $x(t-1)$, shorter time series, and marginally smaller exponents, but similar trends.

In figure~\ref{fig_me}A, we demonstrate the fitting procedure for the 2000 cohort, career age 15, and productivity. The pale points are the observations for authors with $p_i(15)\geq1$ and $P_i(14)\geq1$. The full points result from putting these observations into 20 bins of exponentially increasing size. The model is fitted to the binned data using the method of ordinary least squares, and the coefficient of determination $R^2$ quantifies how well the model fits the corresponding unbinned data. The lower cutoff is estimated by choosing $x_{\mathrm{min}}$ such that $R^2(x_{\mathrm{min}})$ has its first maximum. Such a method that includes the identification of the lower cutoff is not discussed in the literature \citep{perc_matthew_2014}. Ours is a simple heuristic that, in our particular application scenario, underestimates both model parameters but mitigates statistical errors on the scaling exponent as well as biases from finite-size effects.

\paragraph{Independent variables:}
There are substantive and methodological reasons to not mix data from different cohorts. Substantively, we are interested in changes that may have occurred as computer science grew and became a team-based science. Methodologically, it prevents to account for variations in the productivity and impact functions of authors across career ages \citep{penner_predictability_2013}.
To account for this, the models contain a \textit{cohort} variable.

To test for a gender effect, we include a \textit{gender} category. Since gender could not be detected for all authors, we use \textit{male}, \textit{female}, and \textit{undetected} as dummy variables.

We are further interested in the factors that affect an author's career. According to the Matthew Effect, advantages accumulate over time. The earlier in a career an advantage sets in, the more it can accumulate. Hence, all our independent variables are computed for the early career $[1,t_{\mathrm{e}}]$. We choose $t_\mathrm{e}=3$ to delimit the early career. 

The first construct category contains the early-career \textit{achievements} of authors. \textit{Productivity} is the cumulative number of publications $P_i(t_{\mathrm{e}})$. \textit{Productivity (1st author)} $P_{i(\mathrm{1st})}(t_{\mathrm{e}})$ is the number of publications written as a first author. \textit{Impact} is the cumulative number of citations $C_i(t_{\mathrm{e}})$. 
Another way to quantify achievement is to use the reputation of the sources (journals and conference proceedings) an author publishes in. We operationalize symbolic capital based on the $h5$-index \citep{google_google_2020} of sources. $h5_s(y)$ is the maximum cumulative number of publications $h5$ published in source $s$ in the years $[y-4,y]$ that have accumulated at least $h5$ citations in those years. The binary \textit{top source} variable is then 1 if an author has at least one publication in a source that belongs to the top $25\%$ of the distribution in a given year.

Careers are affected by being able to reap benefits from embedding into social networks. Hence, our second construct category is \textit{social support}.
\textit{Collaboration network} is the size of the social support network, measured in terms of the number of distinct co-authors in the early career.
The maturing of the computer science hasfield is marked by the emergence of team science \citep{Wuchty2007}. Therefore, we study the effect of \textit{team size}, defined as the median number of authors of all publications produced in the early career. \textit{Senior support} quantifies the extent to which an author enjoys mentorship from a senior scientist. Our proxy is the largest $h$-index \citep{hirsch_index_2005} of all co-authors $j$ in the social support network: $\mathrm{max}(h_j(y))$. $h_j(y)$ is the maximum cumulative number of publications $h$ that each has accumulated at least $h$ citations until $y$, where $y$ is the year in which author $i$ is in career age $t_{\mathrm{e}}$.

All independent variables are standardized by subtracting the median and dividing the result by the range between the 1st and 3rd quartile.

\paragraph{Dependent variables:}
Authors can leave academia for a certain number of years in a row. We label each author in our corpus as a \textit{dropout} if she or he has not published for ten consecutive years in the first 15 career ages. $59\%$ of the authors are labeled as dropouts. This label is used as a binary variable in dropout predictions.

To quantify the \textit{success} of authors, we define $C_i^\mathrm{+}(15)=C_i(15)-C_i(t_{\mathrm{e}})$. This variable measures citation increase in the cumulative number of \textit{citations} received by all publications published until and in career age $15$ after the early career period. This measure avoids autocorrelation with the independent predictor $C_i(t_{\mathrm{e}})$ and an inflated coefficient of determination \citep{penner_predictability_2013}.

Dependent variables are standardized like the independent ones.

\paragraph{Prediction models:}
In \textit{dropout prediction}, we regress dropout against the independent variables using a logistic model.
In \textit{success prediction}, we regress citation increase against the independent variables using a linear model. We use the elastic net variant since it contains regularization techniques to ensure that the model generalizes well (to avoid overfitting). These techniques estimate weights that penalize regression coefficients. This is useful when multiple independent variables are correlated with each other \citep{hastieElasticNet}. There are two parameters.
The mixing parameter $\lambda$ controls the extent to which overfitting is avoided by L1 regularization (which makes some weights zero, i.e., selects variables to remove) as opposed to L2 regularization (which makes weights small but not zero). When $\lambda=1$ only L1 penalties are applied; when $\lambda=0$ only L2 penalties are applied. We use the default $\lambda=0.5$, that is, the elastic net will perform variable selection but will keep highly correlated variables in the model.
The regularization parameter $\alpha$ is a constant that multiplies the penalty weights. When $\alpha=0$, the model becomes an ordinary-least-squares regression (without any regularization). The optimal value for $\alpha$ is learned from the data.

In all prediction models, there are 292,443 observations. Regression coefficients and their weights are learned in 10-fold cross-validation. That is, the data is randomly divided into 10 folds of 29,244 observations, and in 10 iterations the model is trained on 9 folds and tested on the remaining one \citep{hox_computational_2017}.
Regression coefficients are reported as averages across the 10 folds. When means are far from zero, effects are sizable; when standard deviations are low, coefficients are robust.

For the binary prediction model (dropout prediction), we use two scores as evaluation metrics. The \textit{F1} score is the weighted average of the precision (proportion of predicted positives that are correct) and recall (proportion of known positives that are predicted correctly). The \textit{average precision} summarizes a precision-recall curve as the weighted mean of precisions achieved for every highest value of recall. Both range from 0 to 1.
For the linear models (success prediction), we use two other goodness-of-fit measures. The \textit{mean squared error} quantifies the mean squared distance of all observations to the regression line. The \textit{adjusted $R^2$} coefficient of determination measures the proportion of the variance in the dependent variable that is predictable from the independent variables. It corrects for the number of independent variables that the models use. It increases only if the new term improves the model more than would be expected by chance. Both measures range between 0 and 1 where higher values are better.
For all four evaluation metrics, we report the average value across 10 folds.

\bibliography{biblio.bib}
\appendix
\newpage
\section{Individual inequality for window counting}
\label{appendix}

\begin{figure*}[b!]
    \centering
    All authors \\
    Every Author assignment {} {} {} {} First Author assignment\\
    \includegraphics[scale=0.28]{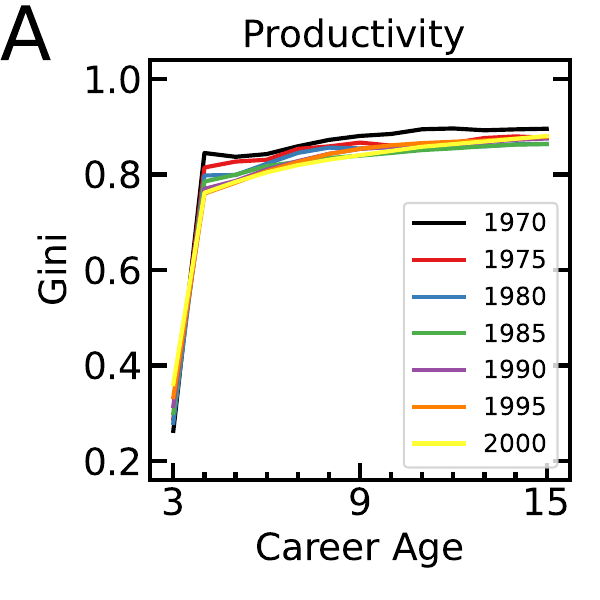}
    \includegraphics[scale=0.28]{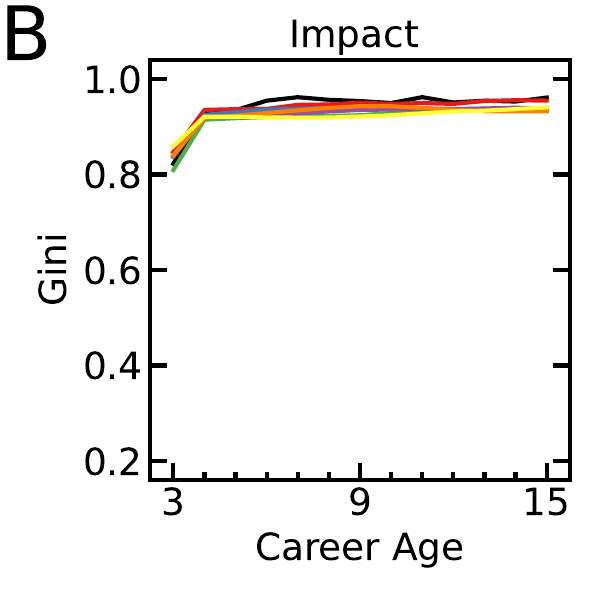}  
    \includegraphics[scale=0.28]{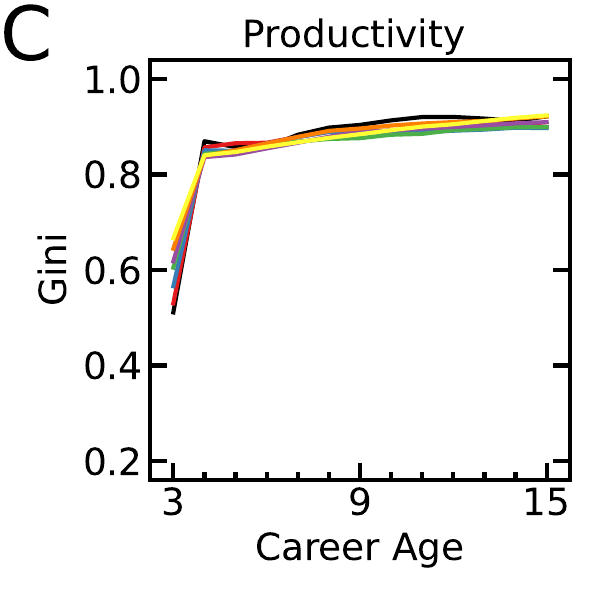}
    \includegraphics[scale=0.28]{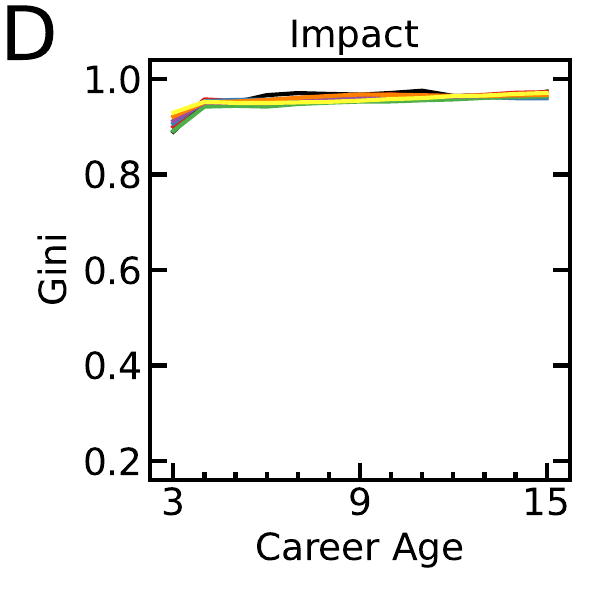}\\
    Dropouts removed\\
    Every Author assignment {} {} {} {} First Author assignment\\
    \includegraphics[scale=0.28]{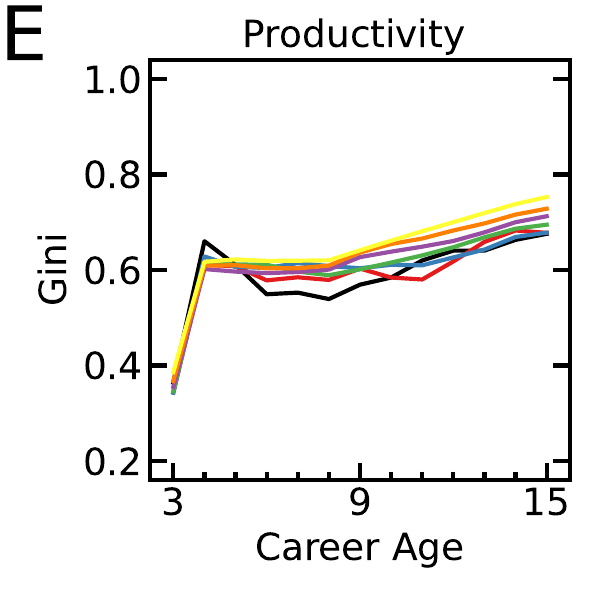}
    \includegraphics[scale=0.28]{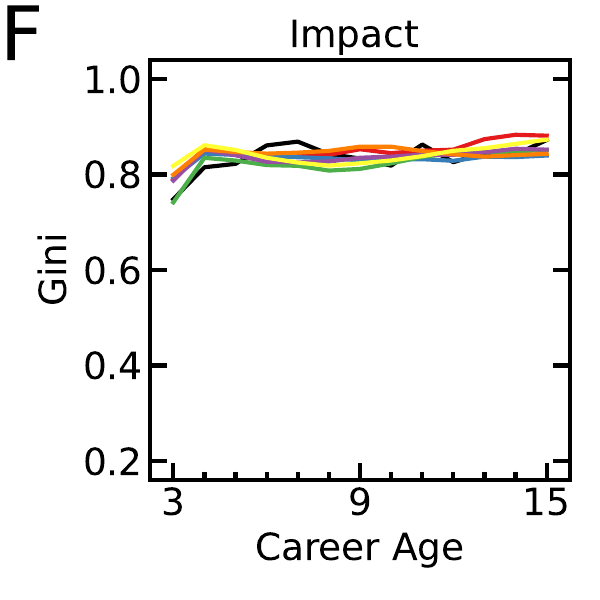}
    \includegraphics[scale=0.28]{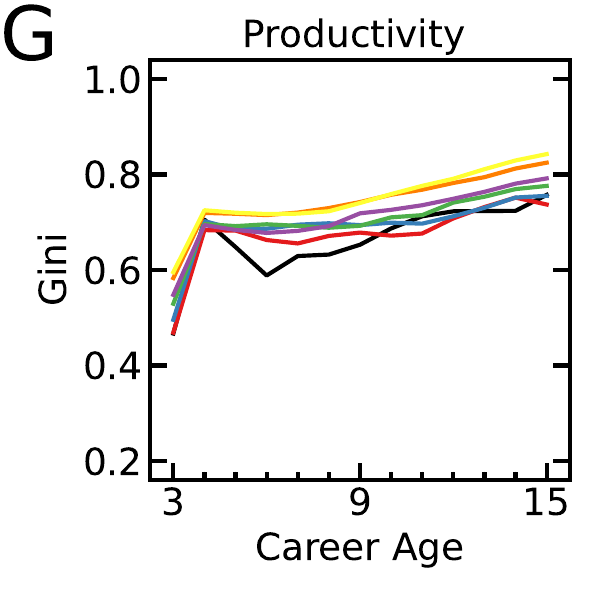}
    \includegraphics[scale=0.28]{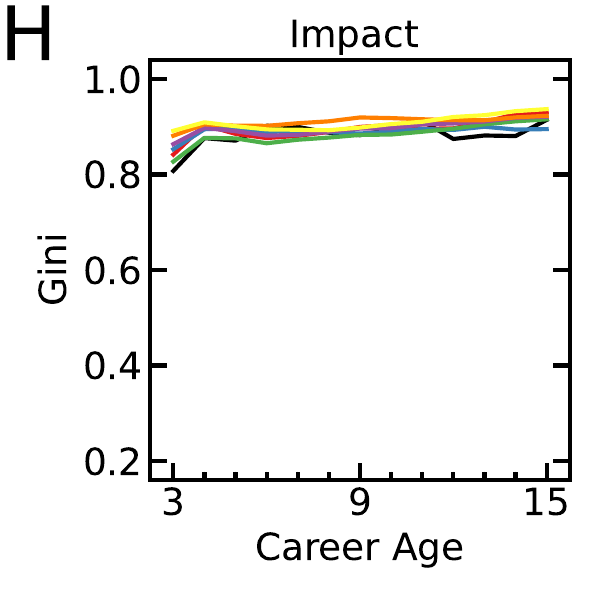}
    \caption{Individual inequality in productivity and impact as a function of career age, depicted for seven cohorts between 1970 and 2000 and estimated using window counting.
    We count publications and citations in 3-year publication windows (given career age plus previous two career ages, $p_{\mathrm{3yr}}(t)$ and $c_{\mathrm{3yr}}(t)$, defined in ``Materials and methods: Individual inequality'').
    (First two columns) Assigning publications to all authors. (Last two columns) Assigning publications only to first authors.
    (Second row) Authors are filtered that have not published for ten consecutive years (most likely left academia).
    Inequality in impact is always larger and more stable over the course of a career than inequality in productivity.
    }
    \label{fig:gini-win}
\end{figure*}

\begin{figure*}
    \centering
    All authors\\
    Every Author assignment {} {} {} {} First Author assignment\\
    \includegraphics[scale=0.28]{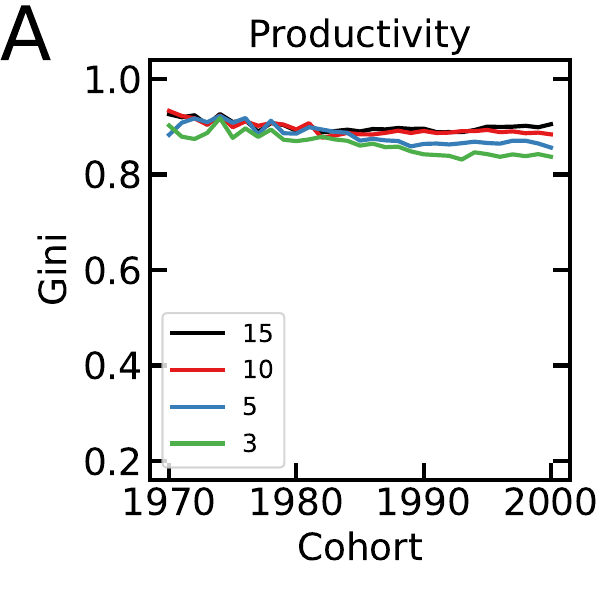}
    \includegraphics[scale=0.28]{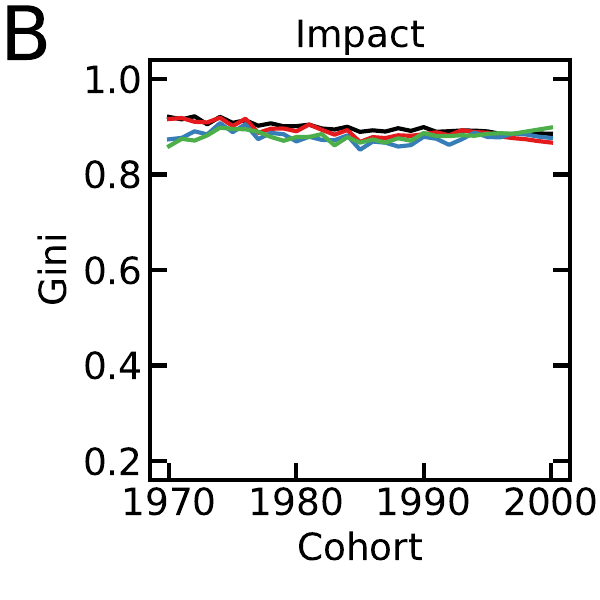}
    \includegraphics[scale=0.28]{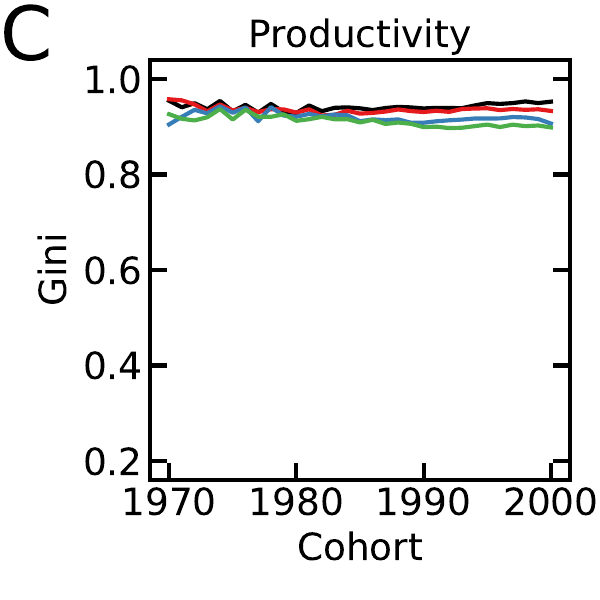}
    \includegraphics[scale=0.28]{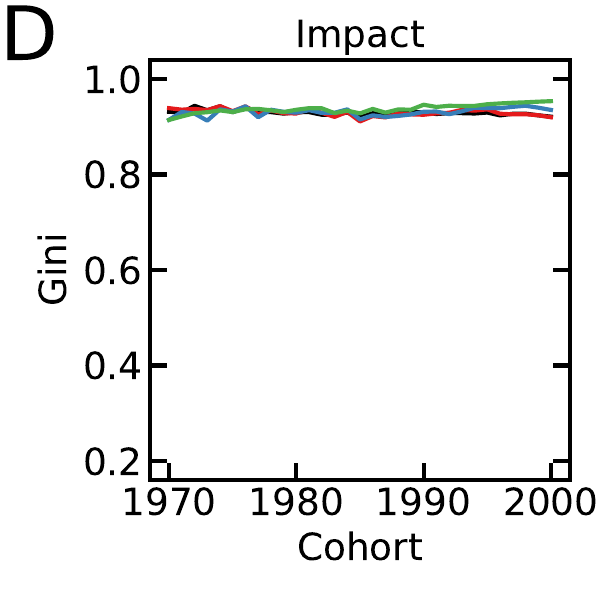}\\
    Dropouts removed\\
    Every Author assignment {} {} {} {} First Author assignment\\
    \includegraphics[scale=0.28]{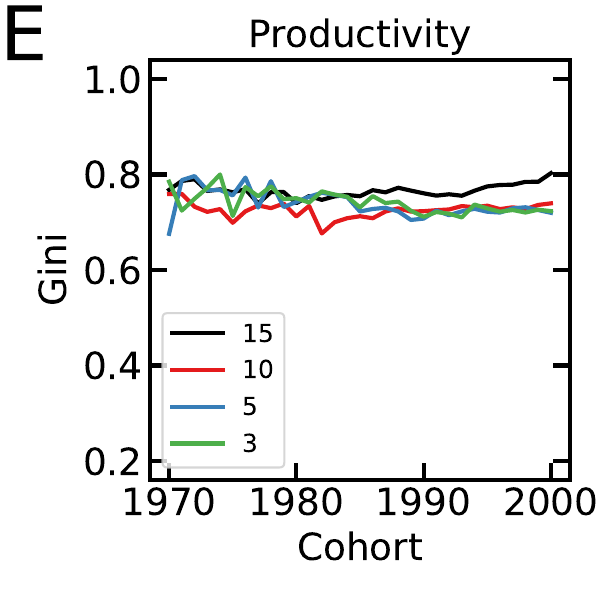}
    \includegraphics[scale=0.28]{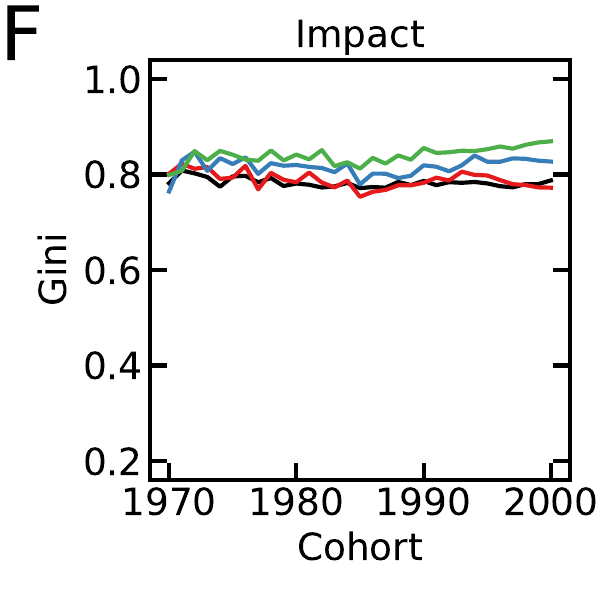}
    \includegraphics[scale=0.28]{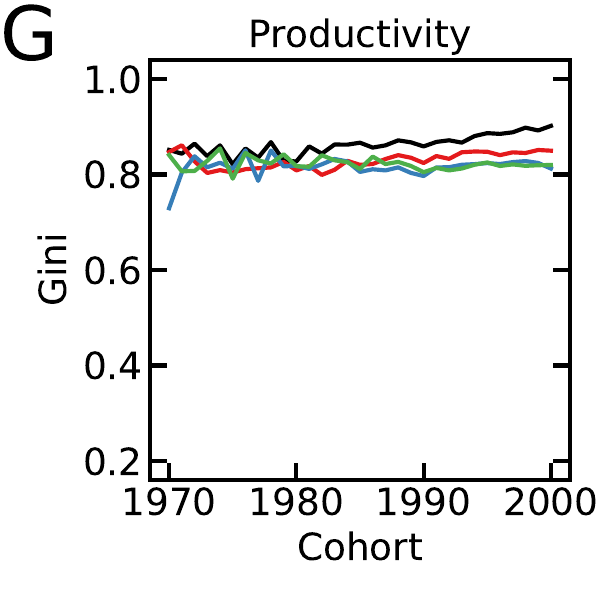}
    \includegraphics[scale=0.28]{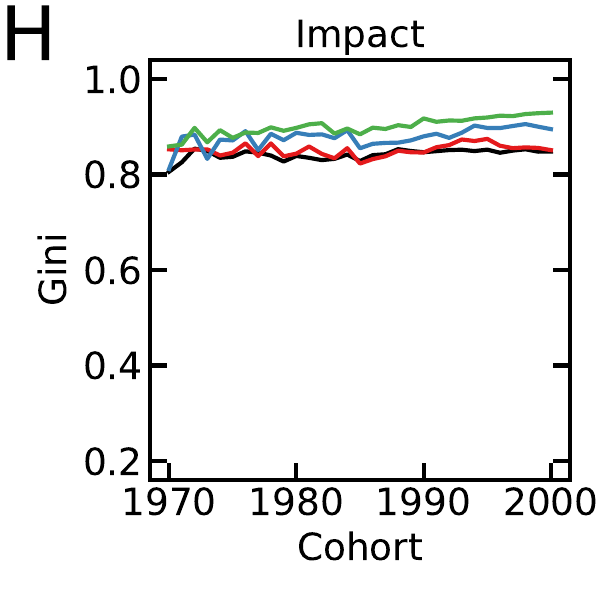}
    \caption{Individual inequality in productivity and impact as a function of cohort start year, depicted for career ages 3, 5, 10, and 15 and estimated using window counting).
    We count the number of publications authored in a career age and the number of citations received in a career age by all publications authored until and in that career age ($p(t)$ and $c(t)$, defined in ``Materials and methods: Individual inequality'').
    (First two columns) Assigning publications to all authors. (Last two columns) Assigning publications only to first authors.
    (Second row) Authors are filtered that have not published for ten consecutive years (most likely left academia).
    Inequality is surprisingly stable over cohorts though they vary in size and the field has evolved over 45 years.
    }
    \label{fig:gini-win-across-cohorts}
\end{figure*}

\end{document}